%% file: main.tex
\documentclass[manuscript,screen,nonacm]{acmart}

\AtBeginDocument{%
  }

\setcopyright{none}

\usepackage{amsmath}
\usepackage{algorithm}
\usepackage{algorithmic}
\usepackage{graphicx}
\usepackage{textcomp}
\usepackage{xcolor}
\usepackage{multirow}
\usepackage[acronym,toc]{glossaries}
\usepackage{enumitem}
\usepackage{makecell}
\usepackage{siunitx}

\begin{document}
% ---------
% Title
% ---------
\title{From Codebase to Culprit (C2C): Reducing the Search Space for Bugs with Semantic Retrieval and Hierarchical Reinforcement Learning}

% Authors
\input{other/authors}%

% ---------
% Abstract
% ---------
\input{sections/abstract}%

% Glossary
\input{other/glossary}%

%% Keywords
\input{other/keywords}%

\maketitle

\input{sections/introduction}%

% -------------
% Related Work
% -------------
\input{sections/related_work}

% -------------
% Dataset
% -------------
\input{sections/dataset}

\input{sections/overview_c2c}

\input{sections/coderetriever}
\input{sections/hrl_localization}

% ----------------------
% Experiments & Results
% ----------------------
\input{sections/results}

% -----------
% Discussion
% -----------
\input{sections/discussion}

% --------------------
% Threats to Validity
% --------------------
\input{sections/threats_to_validity}

% -----------
% Conclusion
% -----------
\input{sections/conclusion}

% -----------
% References
% -----------

\bibliographystyle{ACM-Reference-Format}
\bibliography{references}

\end{document}

%% file: other/authors.tex
\author{Ankur Garg}
\affiliation{%
  \institution{University of Calgary}
  \country{Canada}}
\email{ankur.garg1@ucalgary.ca}

\author{Corey Yang-Smith}
\affiliation{%
  \institution{University of Calgary}
  \country{Canada}}
\email{corey.yangsmith@ucalgary.ca}

\author{Rishav Rishav}
\affiliation{%
  \institution{University of Calgary}
  \country{Canada}}
\email{rishav.rishav@ucalgary.ca}

\author{Ahmad Abdellatif}
\affiliation{%
  \institution{University of Calgary}
  \country{Canada}}
\email{ahmad.abdellatif@ucalgary.ca}

\author{Samira Ebrahimi Kahou}
\affiliation{%
  \institution{University of Calgary}
  \country{Canada}}
\email{samira.ebrahimikahou@ucalgary.ca}

% Author Short List (Headers)
\renewcommand{\shortauthors}{Garg et al.}

%% file: sections/abstract.tex
\begin{abstract}
We introduce C2C (From Codebase to Culprit), a framework for precise bug localization that progressively reduces the debugging search space across multiple levels of granularity: files, functions, and lines of code. To mirror developer's natural top-down debugging workflows, C2C integrates semantic retrieval and Hierarchical Reinforcement Learning (HRL) in a two-stage process. First, it performs recall-oriented retrieval of buggy candidates via semantic vector similarity search using bug-report text, including available stack-trace information, against a database of embeddings, where the embeddings are fine-tuned via contrastive learning with CodeBERT. Building on this reduced search space, the HRL framework incrementally localizes bugs, reasoning from files to functions and ultimately to individual lines of code. Unlike prior approaches which operate at a single granularity, C2C enables multi-resolution localization while maintaining contextual consistency across decisions. Experiments on real-world Java and Python datasets demonstrate that C2C improves retrieval precision and localization accuracy. Ablation studies further highlight the contributions of hierarchical decomposition, structured learning signals, and reward shaping in advancing multi-level bug localization.
\end{abstract}

%% file: other/glossary.tex
\newacronym{rl}{RL}{Reinforcement Learning}
\newacronym{hrl}{HRL}{Hierarchical Reinforcement Learning}
\newacronym{ir}{IR}{Information Retrieval}
\newacronym{mdp}{MDP}{Markov Decision Process}
\newacronym{mrr}{MRR}{Mean Reciprocal Rank}
\newacronym{sbbl}{SBBL}{Spectrum-Based Bug Localization}
\newacronym{llm}{LLM}{Large Language Model}
\newacronym{nn}{NN}{Neural Network}
\newacronym{mbbl}{MBBL}{Mutation-Based Bug Localization}
\newacronym{atlas}{ATLAS}{Adaptive Three-Level Auto-localization System}
\newacronym{rlocator}{RLocator}{Reinforcement Learning Locator}
\newacronym{codebert}{CodeBERT}{Code Bidirectional Encoder Representations from Transformers}
\newacronym{faiss}{FAISS}{Facebook AI Similarity Search}

\newacronym{pt1}{PT1}{Pretraining Agents Sequentially}
\newacronym{pt2}{PT2}{Pretraining with Ground Truths}
\newacronym{vt}{VT}{Vanilla Training}
\newacronym{t1b1}{T1B1}{Training One-by-One}

%% file: other/keywords.tex
\keywords{Bug localization, hierarchical reinforcement learning, software maintenance, code embeddings, vector similarity search}

%% file: sections/introduction.tex
\section{Introduction}
\label{sec:introduction}
Software defects remain among the most significant and costly liabilities in modern IT systems. The Consortium for Information \& Software Quality (CISQ) estimates that legacy-system failures, production outages, security breaches, and accumulated technical debt cost U.S. organizations \textbf{\$2.41 trillion} in 2022 alone~\cite{krasner2022cpsq}. Thus, accurately and efficiently localizing bugs is critical in reducing debugging effort, enhancing software reliability, and managing overall project costs.

However, debugging remains challenging and resource intensive. Research on developer debugging behavior indicates that they frequently employ forward reasoning \cite{BohmeFSE2017} and scientific debugging strategies \cite{perscheid2017studying}, interleaving logical deduction with information gathering and dependency analysis \cite{Ko2006}. Such approaches often require examining extensive code segments, making debugging tedious, error-prone, and costly.

To address these challenges, various bug localization methods have been proposed, including \gls{sbbl} \cite{jones2002visualization, abreu2007accuracy, wong2013dstar}, \gls{ir}-based ranking methods \cite{Zhou_2012_IR_BugLocator, Saha_2013_IR_BLUiR}, and learning-based approaches \cite{Li_2019_LBFL_DeepFL, Chen_2023_CL_SupConFL, Li_2025_CL_FALCON}. While some techniques can be adapted for different granularities (file, function, or line level) \cite{DLFL_Survey}, few natively support localization at multiple granularities simultaneously \cite{ZouJavaFLSurvey, Rezaalipour_Furia_2024}. Addressing this gap is critical, as hierarchical localization augmented with rich context has the potential to outperform single-stage methods, especially when accurate identification at a single level alone proves inadequate.

Recent \gls{rl}-based methods, such as RLocator \cite{Chakraborty_2024_RL_RLocator}, have improved file-level predictions. However, these approaches typically flatten the localization task into single-stage decisions, limiting their capacity for iterative refinement. In practice, software debugging often follows a hierarchical, top-down workflow: developers start from a stack trace, shortlist suspicious files, narrow down to relevant functions, and ultimately identify the precise problematic lines. Inspired by this process, we introduce \textbf{\texttt{C2C}}, a novel \gls{hrl} framework designed to mirror and automate this sequential strategy. 

\texttt{C2C} operates in two stages. It begins with \textbf{CodeRetriever}, a fine-tuned CodeBERT model \cite{codebert} that maps bug-report text, including available stack-trace information, to top-K candidate buggy files using a precomputed vector database. From this candidate pool, the second stage, our HRL architecture, employs three specialized \gls{rl} agents in sequence. Each of these modules is an RL agent in the standard sense: it observes a task-specific state, selects an action from its action space, and is trained with a reward signal to improve bug localization performance \cite{kaelbling1996reinforcement}. Together, these agents progressively refine localization at different levels of granularity, narrowing the search from file, to function, to line. Our three HRL agents are:
\begin{enumerate}
    \item \textbf{File-level agent:} receives the retrieved candidate file set and selects the most likely buggy file.
    \item \textbf{Function-level agent:} receives the selected file and chooses the most likely buggy function within it.
    \item \textbf{Line-level agent:} receives the selected function and identifies the most likely buggy line.
\end{enumerate}

Overall, C2C takes as input a repository and bug-report text obtained from GitHub issues, including available stack-trace and error information. During both training and inference, stack traces are used to progressively narrow the search space from files to functions to lines. Figure~\ref{fig:c2c_overview_fig} provides a high-level overview of our inference pipeline. Sections~\ref{sec:coderetriever} and ~\ref{sec:hrl_localization} then present the motivation and implementation details for CodeRetriever and the HRL architecture, respectively.

We evaluate \texttt{C2C} on two complementary benchmarks: real-world Elasticsearch issues from the BEETLEBOX benchmark \cite{beetlebox} (Java), and 11 Python repositories from SWE-bench Verified \cite{swebench}. 

Across both settings, iterative contrastive fine-tuning with hard negative mining improves retrieval quality over the baseline, achieving a Hit@10 of 57.8\% on Elasticsearch and 69.7\% on SWE-bench (Table~\ref{tab:retrieval_metrics}). Notably, at broader retrieval depths, the model achieves a Hit@50 of 93.8\% on SWE-bench and 72.9\% on Elasticsearch, establishing a solid candidate pool for the downstream localization agents. Building on this retrieval stage, the three-level \gls{rl} agents demonstrate effective fine-grained localization, achieving file/function/line \gls{mrr} of 0.5005/0.5450/0.2932 on Elasticsearch and 0.4156/0.2815/0.1000 on SWE-bench. Ablation studies confirm that iterative contrastive training, hierarchical decomposition, and shaped intermediate rewards each contribute substantially to these results.

We address the following research questions (RQs):
\begin{enumerate}
    \item \textbf{RQ1:} How effectively does CodeRetriever align stack traces and buggy files in the embedding space?
    \item \textbf{RQ2:} How does hierarchical decomposition improve localization compared to single-stage methods?
    \item \textbf{RQ3:} How do different reward strategies affect learning in HRL?
    \item \textbf{RQ4:} Does \texttt{C2C} generalize across programming languages and repositories?
\end{enumerate}

The primary contributions of this paper include:
\begin{enumerate}
    \item Iterative contrastive fine-tuning of CodeBERT with hard-negative mining for effective bug retrieval.
    \item The first open-source \gls{hrl} framework to the best of our knowledge for multi-granular bug localization.
    \item A shaped-reward strategy that stabilizes \gls{hrl} training.
    \item An open-source benchmark, dataset pipeline, and reference implementation to support reproducibility.\footnote{Code and steps to reproduce available at \url{https://github.com/agarg-dev/ATLAS}}
\end{enumerate}

The remainder of our paper is structured as follows: Section~\ref{sec:related-work} discusses related work. In Section~\ref{sec:dataset} we introduce our evaluation benchmarks and discuss how we preprocess the data for our method. We provide an overview of \texttt{C2C} in Section~\ref{sec:overview_c2c}, with detailed explanations on the motivation and implementation of our two stages, CodeRetriever and our HRL Architecture, in Sections~\ref{sec:coderetriever} and~\ref{sec:hrl_localization}, respectively. We then present our experimental results and ablation studies in Section~\ref{sec:results}, discuss our findings in Section~\ref{sec:discussion}, outline threats to validity in Section~\ref{sec:threats}, and conclude in Section~\ref{sec:conclusion}.

%% file: sections/related_work.tex
\section{Related Work}
\label{sec:related-work}
\subsection{Spectrum and Mutation-Based Bug Localization}
\label{subsec:sbfl}

Automated bug localization techniques have evolved from traditional \gls{sbbl} approaches to more advanced methods such as \gls{mbbl} and \gls{ir}-based techniques. 

\gls{sbbl} techniques have served as the foundation of bug localization research by mining contrasts between passing and failing test executions to compute suspiciousness scores and rank buggy code elements \cite{sarhan2022survey, Higor_2016_SBFL_Survey}. Classical methods such as Tarantula \cite{jones2002visualization}, Ochiai \cite{abreu2007accuracy}, and DStar \cite{wong2013dstar} effectively leverage test outcomes, but often struggle in large codebases or in the presence of incomplete coverage and multiple interacting bugs \cite{Wong2016}.

To address these limitations, \gls{mbbl} extends \gls{sbbl} with mutation analysis, where code elements are modified using mutation operators \cite{Papadakis_2012_Mutation}. Metallaxis-FL \cite{Papadakis_2015_Metallaxis} ranks statements higher if mutants are mostly killed by failing tests, while MUSE \cite{Moon_2014_Mutation_MUSE} generates multiple mutant versions of statements covered by failing tests to refine rankings. While \gls{mbbl} reduces reliance on test suites, mutation-based approaches still face scalability challenges in large systems due to the computational cost of mutation analysis \cite{Kim_2021_Mutation}.

\subsection{Information Retrieval and Contrastive Learning-Based Bug Localization}
\label{subsec:irclfl}

As an alternative that reduces the dependency on test coverage, \gls{ir}-based bug localization techniques leverage textual similarity between bug reports and source code elements. For instance, BugLocator \cite{Zhou_2012_IR_BugLocator} uses a revised Vector Space Model to rank candidate files by estimating their similarity to bug reports. BLUiR \cite{Saha_2013_IR_BLUiR} enhances this by treating source code as a structured document, separating constructs like class and method names, and using IR models to match bug reports to relevant files more effectively. Despite these improvements, \gls{ir}-based methods are typically coarse-grained and fail to provide developers with needed contextual information \cite{Wen_2016_IR_Issues}. This limitation motivates the exploration of techniques capable of capturing richer contextual representations.

Contrastive learning methodologies aim to enhance retrieval by improving buggy signal clarity and promoting better generalization across projects. Existing bug localization methods, such as SupConFL \cite{Chen_2023_CL_SupConFL} and FALCON \cite{Li_2025_CL_FALCON}, utilize contrastive techniques to learn embeddings from buggy-fixed code pairs and log traces, respectively.

Single-pass contrastive learning methods using triplet-based loss and hard-negative mining can effectively strengthen semantic representations by identifying challenging negative examples and adaptively refining embeddings. Although such methods have shown promising results in other domains, such as monolithic architecture decomposition \cite{Sellami_2025_CLLLM} and enterprise search \cite{Meghwani_2025_Hard_Negatives}, their application to bug localization remains largely unexplored, representing a valuable direction for future research.

\subsection{Learning and RL-Based Bug Localization}
\label{subsec:lbfl}
Recent research has increasingly leveraged deep learning techniques to improve bug localization by learning representations from code and bug-related data.

One prominent direction focuses on integrating multiple bug localization metrics to construct rich feature sets for learning-to-rank models, such as in MULTRIC \cite{Xuan_2014_LBFL_MULTRIC} and TraPT \cite{Li_2017_LBFL_TraPT}. DeepFL \cite{Li_2019_LBFL_DeepFL} extends this idea the furthest by combining diverse bug localization techniques, complexity-based metrics, and textual similarity to create a comprehensive feature set, significantly outperforming previous methods. Other approaches emphasize improving input representations, such as GRACE's \cite{Lou_2021_LBFL_GRACE} use of Gated Graph Neural Networks to encode structural relationships in code. However, the supervised nature of many of these methods often limits cross-project generalization. 

Recent methods also leverage large pretrained transformers to learn rich code embeddings. FBL-BERT \cite{Ciborowska_2022_LLM_FBL_BERT} analyzes commit histories and uses nearest-neighbor retrieval to identify bug-inducing hunks, while LLMAO \cite{Yang_2024_LLMAO} fine-tunes lightweight bidirectional adapters atop a frozen language model to localize buggy lines of code without relying on test cases. BugCerberus \cite{Chang_2025_LLM_BugCerberus}, the most closely aligned with our work, adopts a hierarchical framework with fine-tuned models at the file, function, and statement levels for precise localization, though its primary focus is automated issue fixing.

While \gls{rl} has been applied to a wide range of software engineering tasks, such as test prioritization \cite{Gupta_2019_RL_Test_Prioritization}, test generation \cite{Chen_2024_RL_Test_Generation, Esnaashari_2021_RL_Test_Gen}, and code generation \cite{Shojaee_2023_RL_Code_Generation}, its application within bug localization remains limited. This is surprising given \gls{rl}'s strength in learning effective search strategies under sparse supervision, especially in settings where explicit bug annotations are scarce or incomplete \cite{Maruyama_2008_RL_Model_FL, Liu_2005_RL_Model_SOBER}. One notable exception is RLocator, which applies an RL framework to localize bugs at the file level \cite{Chakraborty_2024_RL_RLocator}. Another example, SBFL4RL \cite{Moran_2023_RL_SBFL4RL}, combines spectrum-based heuristics with RL agents; however, it is designed for policy learning environments and focuses on identifying bugs in state-space behaviors rather than in source code, limiting its generalizability to traditional bug localization tasks.

\texttt{C2C} addresses these limitations by formulating bug localization as an \gls{hrl} problem \cite{Barto2019}, introducing a structured decision process with separate agents responsible for localization at the file, function, and line levels. Each policy acts on semantically rich states produced by CodeBERT embeddings that we fine-tune using an \emph{iterative, two-stage contrastive learning process}. During testing, downstream agents condition their decisions on the actions chosen upstream, enabling the framework to encompass entire repositories to individual lines without blowing up computational cost. This hierarchical, context-aware design allows \texttt{C2C} to produce multi-granular predictions on large codebases, whereas prior systems such as RLocator and LLMAO are restricted to singly-granular outputs. By combining hierarchical policy learning with contrastively trained contextual code embeddings, \texttt{C2C} provides a unified framework for multi-granular bug localization.

%% file: sections/dataset.tex
\section{Problem Formulation and Datasets}
\label{sec:dataset}
This section describes the benchmarks used in our study, formalizes the multi-granular bug localization task, and explains how we construct the offline artifacts required for retrieval and hierarchical localization.

\subsection{Problem Formulation}
Given a bug report, which may include stack-trace information, the goal of C2C is to localize the corresponding bug at three levels of granularity: the buggy file, the buggy function within that file, and the buggy line within that function.

Rather than predicting over all lines in the repository, we formulate bug localization as a hierarchical sequential decision problem that progressively narrows the search space from files to functions and finally to lines. This process is carried out by multiple reinforcement learning agents, rather than by a single-pass prediction model. A complete prediction is therefore a file-function-line path associated with a bug report. In practice, bugs may span multiple lines, functions, or files in a real software repository. While \texttt{C2C} is trained to identify a single correct target at each hierarchical level, the ranked lists produced by each agent provide natural coverage over multiple locations.

\textbf{Input.}
The input to our framework is a bug report associated with a software repository, as well as the repository itself. We first embed each file (buggy and non-buggy) from the target repository into a vector database which we use downstream for buggy file retrieval. Utilizing the same embedding model, we also extract and embed the associated bug report, represented as a stack trace, into our vector database. The first stage of our framework, CodeRetriever, utilizes the stack trace embedding to retrieve a set of top-K candidate buggy files, which serves as input for the second stage of our approach, the \gls{hrl} architecture of our framework. This two-stage inference process is illustrated in Figure~\ref{fig:c2c_overview_fig}.

\textbf{Output.} The output of \texttt{C2C} is a hierarchical localization prediction consisting of a file, a function within that file, and a line within that function. Together, these form a complete localization path identifying the predicted buggy code element. At inference time, the system produces a structured prediction of the form \textbf{(file, function, line)}, where the predicted line is the final fine-grained bug localization result for the input bug report.

\textbf{Unit of Analysis.}
Each training and evaluation example is a labeled file-function-line instance associated with a bug report, which we extract directly from GitHub issues of the associated repository. Formally, each instance consists of five elements: an issue ID, its stack trace, a buggy file, a buggy function, and a buggy line. A single GitHub issue may therefore produce multiple instances when the corresponding fix affects multiple buggy locations. We further explain our data processing in Section~\ref{subsec:instance-construction}.

\subsection{Datasets}
\label{subsec:datasets}
We conduct our study on two complementary benchmarks spanning Java and Python, summarized in Table~\ref{tab:dataset-comparison}.

\textbf{BEETLEBOX Elasticsearch.}
We use the \textbf{Elasticsearch repository} from the BEETLEBOX benchmark~\cite{beetlebox} as our primary Java dataset. It contains authentic bug reports and code fixes derived from GitHub issues and is enriched with the detailed metadata, stack traces, and annotated buggy locations necessary to train and evaluate \texttt{C2C}. The dataset is comprised of 326 bug reports involving 691 buggy files and 5{,}326 non-buggy files (Table ~\ref{tab:dataset-comparison}). From these 326 bug reports, our preprocessing pipeline (Section~\ref{subsec:instance-construction}) extracts 5{,}342 file-function-line triplets which we use for training and evaluation.

\textbf{SWE-bench Verified.}
To assess the generalizability of \texttt{C2C} across programming languages and repositories, we evaluate \texttt{C2C} on \textbf{SWE-bench Verified}~\cite{swebench}, a curated subset of 500 high-quality software issues drawn from the larger SWE-bench benchmark and verified by humans. SWE-bench Verified spans \textbf{11 Python repositories}: astropy, django, matplotlib, pylint, pytest, requests, scikit-learn, seaborn, sphinx, sympy, and xarray. From these 500 issues, our preprocessing pipeline (Section~\ref{subsec:instance-construction}) extracts 1{,}972 file-function-line triplets across 393 unique issues (Table~\ref{tab:dataset-comparison}). This multi-repository Python setting complements the single-repository Java evaluation on Elasticsearch and directly addresses the question of whether \texttt{C2C}'s approach generalizes beyond a single project and language.

While these benchmarks provide raw issues, stack traces, and revision metadata, they do not directly provide the structured inputs required by \texttt{C2C}. We therefore apply a preprocessing pipeline to transform each issue into file-function-line instances and the offline artifacts needed for retrieval and hierarchical localization. We describe this pipeline next.

\input{tables/dataset_statistics}

\subsection{Instance Construction}
\label{subsec:instance-construction}
Although BEETLEBOX and SWE-bench Verified provide the raw ingredients for bug localization, they must be further processed before they can be used to train and evaluate \texttt{C2C}. We therefore construct a preprocessing pipeline that converts each issue into structured file-function-line instances and the auxiliary artifacts required by both CodeRetriever and the \gls{hrl} framework. The pipeline has three stages: (1) metadata and code extraction, (2) triplet construction for retriever training, and (3) HRL dataset construction. The output of each stage feeds into the next, as illustrated in Figure~\ref{fig:data_preprocessing}.

To support efficient nearest-neighbor search during retrieval, we additionally index embedded source files using \gls{faiss}, a vector similarity search library used in our retrieval stage.

\input{figures/data_processing_workflow}

\subsubsection{Metadata and Code Extraction}
\label{metadata-code-extraction}
For each issue, we first recover the metadata needed to reconstruct the debugging context and the ground-truth buggy location. Specifically, we extract:
\begin{itemize}
    \item \textbf{Stack Trace Information:} Obtained from the bug report text, including available stack-trace and error context.
    \item \textbf{Commit Identifiers}: Before-fix and latest (HEAD) commit SHAs to reconstruct repository states.
    \item \textbf{Precise Bug Locations}: Derived from commit diffs to isolate the affected file names, function names, and line numbers.
    \item \textbf{Buggy Code Files}: Retrieved from the pre-fix commit to capture genuine buggy code.
    \item \textbf{Non-Buggy Files}: Sampled from the repository at the HEAD SHA, excluding the original buggy file. We examine commit history to account for any file name changes over time.
\end{itemize}

Each data point in the processed dataset corresponds to a unique file-function-line triplet, so multiple entries may share the same issue ID or stack trace. To prevent false negatives during contrastive training, pairs sharing the same issue ID or the same positive file path are masked from the in-batch loss denominator; see Section~\ref{subsec:coderetriever} for details.

\subsubsection{Triplet Construction for Contrastive Retrieval Training}
To train the retrieval component, we convert the extracted artifacts into contrastive triplets. Each triplet consists of a stack trace as the anchor, the corresponding buggy file as the positive example, and a non-buggy file as the negative example. The first round of contrastive learning uses randomly sampled negatives. After this initial training, we embed all files into a \gls{faiss} index and perform hard-negative mining to identify semantically similar non-buggy files. These refined triplets are then used in a second fine-tuning phase, yielding embeddings that support recall-oriented retrieval of buggy files. Further details on the contrastive learning process are provided in Section~\ref{subsec:coderetriever}.

\subsubsection{HRL Dataset Construction}
Using the final retriever, we construct the dataset used for hierarchical localization. For each labeled instance, we encode the stack trace, retrieve a ranked set of top-$K$ candidate files, and record the position of the ground-truth buggy file within that candidate list. If the ground-truth buggy file is not among the top-\(K\) candidates, it is inserted into the candidate set to ensure that the HRL agents always have access to the correct target during training and evaluation. We additionally record whether the ground-truth file was originally retrieved by CodeRetriever, whose retrieval performance is evaluated separately on the unmodified rankings.

Each HRL instance contains the following fields:
\begin{itemize}
    \item \textbf{Issue ID}: The bug report identifier.
    \item \textbf{Stack Trace Embedding}: Encoded using the fine-tuned CodeBERT model.
    \item \textbf{Candidate Files List}: A fixed-size set of $K$ candidate files, initialized from CodeRetriever’s retrieval results and augmented with the ground-truth file when it is not originally retrieved.
    \item \textbf{Buggy File Index}: The position of the buggy file in the candidate list.
    \item \textbf{Buggy Function Name}: The buggy method or function within the buggy file.
    \item \textbf{Buggy Line Number}: The specific line of code containing the bug.
\end{itemize}

For issues with multiple buggy files, a separate HRL instance is constructed for each buggy file. At the file level, candidate positions associated with the same issue are treated as valid targets during training and evaluation.

For Elasticsearch, a single \gls{faiss} index covers the full repository, while for SWE-bench, separate per-repository indices are built. Index construction details are given in Section~\ref{sec:coderetriever}.

\subsection{Data Splits and Evaluation Protocol}
\label{subsec:data-splits}
Both stages of \texttt{C2C}---CodeRetriever and the HRL agents---share a single, coordinated split to prevent data leakage between training and evaluation.

All splits are performed randomly at the \textbf{issue level}: issues (not individual triplets) are assigned to train, validation, or test partitions using an 80/10/10 ratio, \textbf{stratified by repository} to ensure each split contains a representative distribution of projects. This prevents data leakage through shared stack traces, since all triplets derived from the same issue are guaranteed to appear in the same partition. For HRL training and evaluation, the candidate sets are constructed as described in Section 3.3.3, ensuring that the ground-truth file is available to the downstream agents; CodeRetriever itself is evaluated separately using its original retrieval rankings.

For the Elasticsearch dataset, 326 issues yield 5{,}342 triplets, split as: 260 train / 33 validation / 33 test issues (3{,}592 / 1{,}278 / 472 triplets respectively). For SWE-bench Verified, 393 issues yield 1{,}972 triplets, split as: 314 train / 39 validation / 40 test issues (1{,}641 / 186 / 145 triplets respectively). During contrastive training, test-split triplets are excluded entirely.

%% file: tables/dataset_statistics.tex
\begin{table}[t]
  \centering
  \caption{Summary statistics for the Elasticsearch and SWE-bench Verified datasets.}
  \label{tab:dataset-comparison}
  \small
  \setlength{\tabcolsep}{4.5pt}
  \begin{tabular}{@{}lcc@{}}
    \toprule
    \textbf{Characteristic} & \textbf{Elasticsearch} & \textbf{SWE-bench Verified} \\
    \midrule

    \addlinespace[3pt]
    \multicolumn{3}{@{}l}{\textbf{Dataset scale}} \\
    \addlinespace[1pt]
    Repositories & 1 & 11 \\
    Bug reports & 326 & 393 \\
    Bug-location triplets & 5,342 & 1,972 \\
    Unique buggy files & 691 & 477 \\
    Unique non-buggy files & 5,326 & 1,929 \\

    \addlinespace[4pt]
    \midrule
    \addlinespace[2pt]
    \multicolumn{3}{@{}l}{\textbf{Bug scope and dispersion}} \\
    \addlinespace[1pt]
    Multi-file bug reports & 167 & 51 \\
    Multi-function bug reports & 168 & 86 \\
    Multi-line bug reports & 247 & 242 \\
    Median buggy files per report & 2 & 1 \\
    Median buggy lines per report & 5 & 2 \\
    Mean triplets per report & 16.39 & 5.02 \\

    \addlinespace[4pt]
    \midrule
    \addlinespace[2pt]
    \multicolumn{3}{@{}l}{\textbf{Buggy-file complexity}} \\
    \addlinespace[1pt]
    Median lines per buggy file & 327 & 670 \\
    Median functions per buggy file & 13 & 38 \\
    Median LOC per function in buggy files & 7 & 8 \\
    \bottomrule
  \end{tabular}
\end{table}

%% file: figures/data_processing_workflow.tex
\begin{figure}[t]
  \centering
  \includegraphics[width=\textwidth]{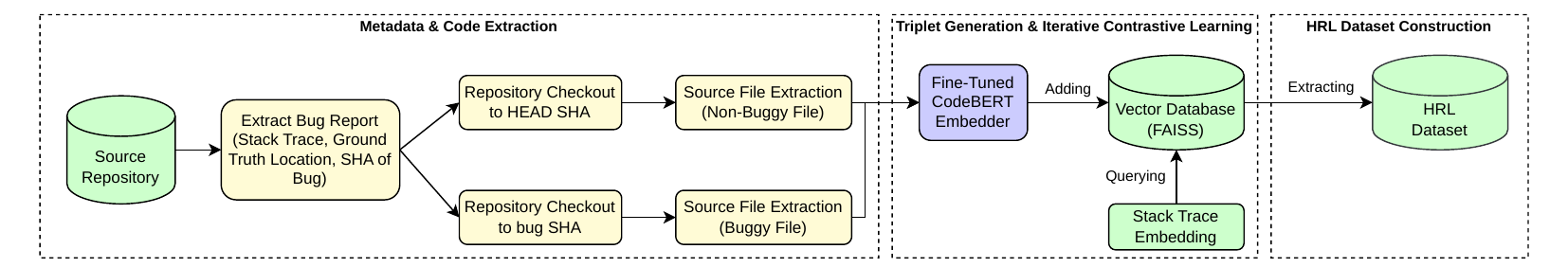}
  \caption{Data extraction and preprocessing pipeline for \texttt{C2C}. The system ingests bug reports and ground-truth data from a source repository, checks out buggy and HEAD SHAs, and extracts buggy files (from buggy SHA), all non-buggy files (from HEAD SHA), and metadata (e.g., issue ID). These form triplets for contrastive training of CodeBERT. Embeddings are stored in a FAISS vector index and refined iteratively via hard negative mining to produce a hierarchical dataset for top-$K$ buggy file retrieval.}
  \Description{A pipeline starting from BEETLEBOX and SWE-Bench Verified bug reports and ground truth, checking out buggy and HEAD SHAs, extracting buggy (from buggy SHA) and non-buggy files (from HEAD SHA) plus metadata, forming triplets for contrastive training of CodeBERT, then storing embeddings in a FAISS index that is iteratively refined via hard negative mining to generate the final hierarchical dataset.}
  \label{fig:data_preprocessing}
\end{figure}

%% file: sections/overview_c2c.tex
\section{Overview of \texttt{C2C}}
\input{figures/c2c_overview}

\label{sec:overview_c2c}
To address the hierarchical nature of bug localization, we decompose the bug localization task into two stages: (1) \textbf{candidate file retrieval}, and (2) \textbf{fine-grained localization across multiple levels of granularity}. The first stage, \texttt{CodeRetriever}, is a semantic retrieval module that fine-tunes CodeBERT using contrastive learning and leverages a \gls{faiss}-based vector database to retrieve a compact set of $K$ candidate files for a given bug report. The second stage, \texttt{C2C}’s \gls{hrl} framework, receives these candidates and performs multi-granular localization at the file, function, and line levels.

This design models bug localization as a sequence of interdependent decisions, where each level refines the search space and provides context for subsequent levels.

\subsection{End-to-End Method Inference}
At inference time, \texttt{C2C} performs bug localization in two stages. First, CodeRetriever maps the input bug report, represented primarily by its stack trace embedding, to a ranked set of candidate source files. Second, a hierarchy of three localization agents progressively narrows this candidate space by selecting the most likely buggy file, the most likely buggy function within that file, and finally the most likely buggy line within that function.

This design separates recall-oriented retrieval from precision-oriented localization. The retrieval stage reduces the repository-level search space to a compact candidate set, while the hierarchical localization stage refines the prediction across increasingly fine levels of granularity. The resulting output predicts a file-function-line localization path associated with the input bug report.

The retriever and the hierarchical agents serve complementary roles. CodeRetriever is optimized to retrieve relevant files, whereas the downstream HRL framework is optimized to rank the correct location as highly as possible within its candidate search space. A summary of our two-stage inference pipeline is available in Figure~\ref{fig:c2c_overview_fig}.

\subsection{Training Overview}
Training is split into two levels based on our two components. CodeRetriever is trained iteratively using contrastive triplets we construct. We utilize iterative negative mining for contrastive learning to improve the model's ability to distinguish between semantically similar code files. Considering the input into the HRL pipeline is a set of candidate files, we utilize our then-trained CodeRetriever to retrieve buggy candidate files for a given bug report and repository. CodeRetriever produces a candidate file pool which we utilize downstream to train our HRL framework. If CodeRetriever fails to retrieve the correct buggy file for a given sample, we inject the ground-truth file into the fixed-size candidate set, ensuring that the HRL framework has a valid target to learn from. We further explain in detail the training setup and protocols for CodeRetriever and the \gls{hrl} framework in Section~\ref{sec:coderetriever} and Section~\ref{sec:hrl_localization}, respectively.

%% file: figures/c2c_overview.tex
\begin{figure}[t]
  \centering
  \captionsetup{skip=4pt}
  \includegraphics[width=\textwidth]{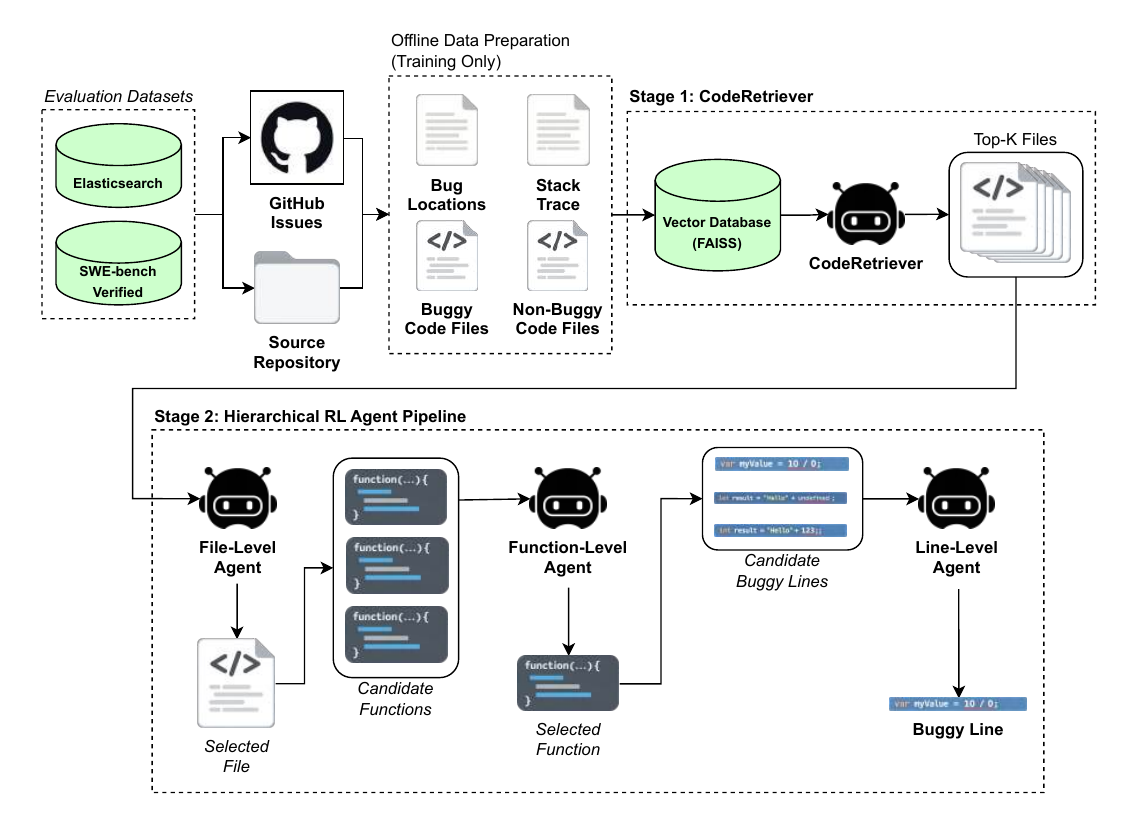}
  \caption{C2C's end-to-end inference pipeline from bug report to final buggy line prediction.}
  \label{fig:c2c_overview_fig}
\end{figure}

%% file: sections/coderetriever.tex
\section{CodeRetriever}
\label{sec:coderetriever}

This section describes CodeRetriever, the retrieval component that forms the first stage of \texttt{C2C}. Its goal is to reduce the repository-level search space by mapping a bug report to a ranked set of candidate files that are likely to contain the bug. We first describe the design and motivation of the retrieval module, then the iterative contrastive training procedure and the construction of the vector index.

\subsection{Overview and Design Rationale}
\label{subsec:coderetriever}

CodeRetriever is a semantic retrieval module based on a fine-tuned CodeBERT encoder and a \gls{faiss} vector database. During training, CodeBERT is fine-tuned via iterative contrastive learning with hard negative mining to align stack traces and code files in a shared embedding space.

We selected CodeBERT as the backbone of CodeRetriever because our first-stage task is fundamentally a cross-modal retrieval problem: the model must map bug reports, represented primarily as stack traces, and source-code files into a shared embedding space for nearest-neighbor search. CodeBERT is a bidirectional encoder pretrained jointly on natural language and code, making it a strong fit for this setting. In contrast to decoder-only code models, which are primarily optimized for generation, our retrieval stage depends on fixed-dimensional embeddings that can be indexed efficiently in \gls{faiss} and compared at scale across entire repositories. We therefore chose CodeBERT not as a claim that it is the universally best code model for bug localization, but because its encoder architecture, joint NL-code pretraining, and computational efficiency align with the requirements of dense semantic retrieval in \texttt{C2C}.

\subsection{Iterative Contrastive Fine-Tuning}  
During the training phase, we fine-tune CodeBERT to produce semantically meaningful embeddings for both stack traces and code files. This fine-tuning uses a hybrid loss function that combines (1) in-batch contrastive loss and (2) margin-based triplet loss, aligning stack trace embeddings with corresponding buggy file embeddings in a shared latent space. 

The process begins with the construction of initial triplets $(a, p, n)$, where $a$ (anchor) is the stack trace embedding, $p$ (positive) is the correct buggy file embedding, and $n$ (negative) is a randomly sampled non-buggy file from the repository at HEAD SHA. The in-batch loss encourages similarity between anchors and their corresponding positives while discouraging similarity to valid in-batch negatives, scaled by temperature $\tau$. To prevent false negatives, we exclude from the denominator any pair $(i, j)$ where $j \neq i$ but triplet $j$ shares the same issue ID as $i$, or whose positive file has the same path as $p_i$ — since such pairs carry nearly identical embeddings despite being distinct training examples. Let $\mathcal{V}(i)$ denote the valid index set for anchor $i$: all $j$ such that $j = i$, or both issue$(j) \neq \text{issue}(i)$ and $\text{path}(p_j) \neq \text{path}(p_i)$. The loss is:

\[
\mathcal{L}_{\text{in-batch}} = -\frac{1}{B} \sum_{i=1}^{B} \log \frac{\exp\!\left(\text{sim}(a_i, p_i) / \tau\right)}{\sum_{j \in \mathcal{V}(i)} \exp\!\left(\text{sim}(a_i, p_j) / \tau\right)},
\]
where $\text{sim}(\cdot,\cdot)$ denotes dot-product similarity and $\tau = 0.1$ is the temperature. We additionally compute $\mathcal{L}_{\text{in-batch}}$ in the symmetric direction (positive-to-anchor) and average both directions to improve embedding consistency.

The margin-based loss explicitly pushes anchor-positive pairs closer while separating anchor-negative pairs by a margin $m$:
\[
\mathcal{L}_{\text{margin}} = \frac{1}{B} \sum_{i=1}^{B} \max\left(0, m - \text{sim}(a_i, p_i) + \text{sim}(a_i, n_i)\right),
\]
where $m = 0.2$.

These two losses are combined into a hybrid objective:  
\[
\mathcal{L}_{\text{hybrid}} = \mathcal{L}_{\text{in-batch}} + \mathcal{L}_{\text{margin}}.
\]

To enhance the embedding space, we adopt \textbf{hard negative mining}. After the initial fine-tuning, we embed the entire corpus and index it in \gls{faiss}. For each buggy file, we identify semantically similar non-buggy files from the \gls{faiss} index to serve as hard negatives. These hard negatives are then used in a second fine-tuning phase, allowing CodeBERT to more effectively distinguish subtle differences between buggy and non-buggy files. Figure~\ref{fig:codebert_finetune} illustrates the iterative contrastive fine-tuning process for CodeBERT.

\input{figures/codebert_finetune}

At inference, CodeRetriever leverages the fine-tuned CodeBERT and \gls{faiss} index to encode the stack trace and retrieve the top-$K$ most relevant candidate files, which are then passed to the \gls{hrl} agents for multi-granular localization.

\subsection{Vector Database Construction}

To support scalable and efficient similarity search over large codebases, we build a \gls{faiss}-based vector database. Each source file is encoded into a 768-dimensional vector using CodeBERT, and indexed with cosine similarity for fast nearest-neighbor queries.

For both the \textbf{Elasticsearch} and \textbf{SWE-bench} datasets, we establish the base corpus by indexing all source files from the latest repository SHA. To accurately represent the bug context, we then inject the ground-truth buggy files extracted from their respective pre-fix SHAs. To prevent data duplication and ensure the retriever targets the correct historical state, we explicitly exclude the latest-SHA versions of those specific buggy files from the index. By indexing the full repositories in this manner, we preserve the extreme natural class imbalance found in real-world software development, providing a highly realistic retrieval challenge.

We build one \gls{faiss} index per repository: a single index for Elasticsearch (one repository) and separate indices for SWE-bench (11 repositories), which avoids conflating embeddings across projects. Before embedding, source files are preprocessed for both datasets, details for which are mentioned in Section ~\ref{subsec:coderetriever-impl}

During training, we first construct the vector database using embeddings from CodeBERT fine-tuned on randomly sampled triplets. We then perform hard negative mining by retrieving non-buggy files that are semantically similar to buggy ones, forming a more challenging triplet dataset for a second round of fine-tuning. The final \gls{faiss} index is built by re-encoding the original corpus using the updated CodeBERT model.

At inference, CodeRetriever encodes the input stack trace and retrieves the top-$K$ most relevant files using this final index. To minimize storage overhead, function and line-level embeddings are generated on demand during downstream localization.

\subsection{Implementation Details}
\label{subsec:coderetriever-impl}
To address CodeBERT's 512-token context limit, we apply a sliding window of 512 tokens (stride 510) to split large files into chunks, embedding up to 50 chunks per file with the final representation taken as the mean of chunk embeddings. Before embedding, source files are preprocessed to reduce noise. Python files have docstrings removed via AST analysis and comments stripped; Java files have block and line comments, import statements, package declarations, and blank lines removed. Preprocessing is applied to code files only; stack traces are embedded as-is. For file-level embeddings, we also include the repository-relative file path along with the source-code content.

Both fine-tuning rounds use the AdamW optimizer with an effective batch size of 64. In the first round (random negatives), the learning rate is $2\times10^{-5}$ with a linear warmup over 10\% of training steps and early stopping with patience 3 over a maximum of 10 epochs. The second round (hard negatives) warm-starts from the first-round checkpoint with a reduced learning rate of $1\times10^{-5}$, no warmup, patience 5, and up to 15 epochs. The best checkpoint is selected by validation loss in both rounds.

%% file: figures/codebert_finetune.tex
\begin{figure}[t]
  \centering
  \includegraphics[width=0.7\columnwidth]{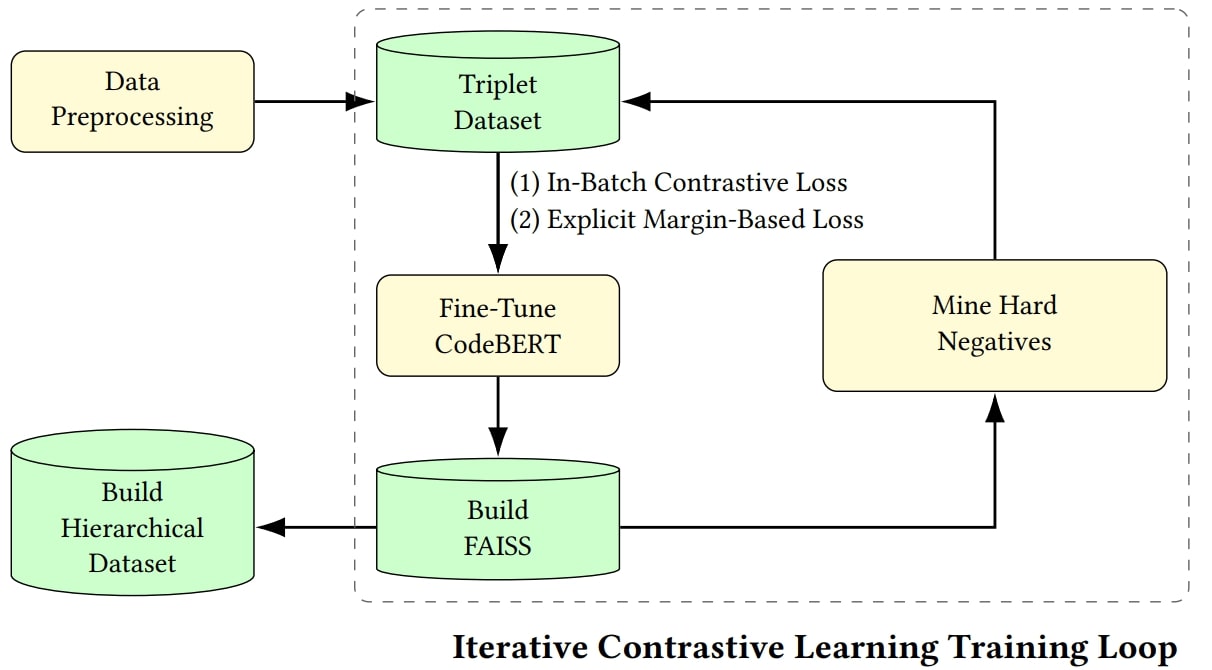}
  \caption{Iterative CodeBERT fine-tuning pipeline in \texttt{C2C}. Raw bug reports and source code are distilled into triplets, then used to fine-tune CodeBERT. After each training pass, a FAISS index mines hard negatives that feed back into the next retraining cycle. The final FAISS index produces the hierarchical evaluation dataset.}
  \Description{A pipeline showing: input bug reports and source code → triplet distillation → CodeBERT fine-tuning → FAISS hard-negative mining → retraining loop → final FAISS index for evaluation.}
  \label{fig:codebert_finetune}
\end{figure}

%% file: sections/hrl_localization.tex
\section{HRL Localization}
\label{sec:hrl_localization}

After CodeRetriever narrows the repository-level search space, \texttt{C2C} performs fine-grained localization through a hierarchy of reinforcement learning agents. This section describes how bug localization is formulated as a sequential decision process over files, functions, and lines, and explains the design of the hierarchical architecture, the reward mechanism and the training strategy.

\subsection{HRL Formulation of the Task}
\label{subsec:problem-formulation}
We model the hierarchical bug localization task as an \gls{hrl} problem, characterized by a sequence of inter-related \gls{rl} agents making progressively finer-grained decisions to pinpoint the exact file, function, and line containing a bug. Each agent operates as an \gls{rl} policy within a \gls{mdp}, wherein at each hierarchical level an agent observes a state $s$, selects an action $a$ from a learned policy distribution $\pi(a|s)$, and subsequently receives a scalar reward $r$ based on the accuracy of its prediction.

Formally, an \gls{mdp} is represented as a tuple $(\mathcal{S}, \mathcal{A}, P, R, \gamma)$, where $\mathcal{S}$ denotes the state space, $\mathcal{A}$ the action space, $P$ the transition dynamics, $R$ the reward function, and $\gamma$ the discount factor. In our methodology, the state space at each level integrates a bug report embedding $s_{\text{bug}} \in \mathbb{R}^{768}$ generated via CodeBERT, alongside embeddings of candidate code entities (files, functions, or lines). The action space involves selecting one candidate from a discrete set at each level of granularity. Rewards are either intermediate (provided at each decision step) or sparse (given exclusively at the final prediction stage). Due to the hierarchical structure, each agent's decision and subsequent state representation depend explicitly on the outcomes of higher-level agents.

\subsection{HRL Architecture}
\label{subsec:astlas-architecture}

Our bug localization framework is structured as a hierarchical reinforcement learning (HRL) architecture consisting of three specialized agents, each operating at a different level of granularity: file, function, and line.

\textbf{File-Level Agent.} The hierarchy begins with the file-level agent, which is responsible for identifying the most probable buggy file. This agent receives $N$ candidate files retrieved through semantic search based on the bug report. The agent's policy, $\pi_f(a_f|s_f)$, processes each file by first creating a unified representation. This is achieved by concatenating the bug report's embedding with the embedding of the candidate file. This combined representation is then fed into a feed-forward network to generate a relevance score for the file. These scores are subsequently normalized to create a probability distribution over the $N$ candidate files. The action, $a_f$, corresponds to selecting a file index based on this distribution.

\textbf{Function-Level Agent.} Once the file-level agent predicts a file, the function-level agent is activated to pinpoint the specific function containing the bug within that file. Its policy, $\pi_{\text{func}}(a_{\text{func}}|s_{\text{func}})$, evaluates $M$ candidate functions that have been extracted from the selected file using static analysis (AST parsing for Python; regex-based brace matching for Java). Similar to the file-level agent, it computes a relevance score for each function by concatenating the bug report embedding with the candidate function's CodeBERT embedding. These scores are then used to form a probability distribution over the function indices. To ensure the agent is trained effectively, it is only activated and supervised when the file predicted by the higher-level agent is correct. For bugs that are not confined to a specific function, a synthetic "global function" is created to represent the file's global scope, ensuring all code is considered.

\textbf{Line-Level Agent.} At the lowest level of the hierarchy, the line-level agent performs fine-grained localization by identifying the exact buggy line within the function selected by the function-level agent. Given $T$ candidate lines from the predicted function, the agent's policy, $\pi_\ell(a_\ell|s_\ell)$, computes a probability distribution over the line indices. It follows the same scoring mechanism as the other agents, creating a combined representation from the line's embedding and the bug report's embedding. The activation of this agent is contingent upon the correct identification of both the file and the function, providing it with a highly relevant context for precise localization.

Architecturally, the file-level and line-level agents each employ a \textbf{Bidirectional LSTM} (BiLSTM) encoder to capture sequential dependencies among candidate embeddings, while the function-level agent uses an MLP encoder. The line-level agent additionally incorporates sinusoidal positional encoding to preserve line-order information. All agents share the same downstream structure: a feature-fusion MLP that concatenates the bug report embedding with the encoded candidate representation, followed by separate ranking and confidence heads.

The overall architecture is depicted in Figure~\ref{fig:atlas_arch}. Training details are provided in Section~\ref{subsec:training}.

\input{figures/architecture_diagram}

\subsection{Reward Design}
\label{subsec:reward-design}

To effectively train our hierarchical agents, we designed a reward system that provides immediate and informative feedback at each level of the decision-making process. The core of our approach is an \textbf{intermediate reward} mechanism that assigns credit based on the accuracy of each agent's predictions, rather than waiting for the final outcome.

\paragraph{Position-Based Intermediate Rewards.}
The reward for each agent is determined by the rank of the ground-truth item within its predicted list. We define a reward vector $W = [w_1, w_2, \dots, w_K]$ where the weights are positive and monotonically decreasing ($w_1 > w_2 > \dots > w_K > 0$). If an agent places the correct item at position $p$ in its top-K ranking, it receives a reward of $r = w_p$. If the correct item is not within the top-K predictions, the reward is zero. This structure incentivizes the agents to rank the correct item as highly as possible.

This position-based reward, $r$, is applied conditionally at each level of the hierarchy:

\begin{itemize}
    \item \textbf{File-Level Reward ($r_{\text{file}}$):} The file-level agent receives a reward, $r_{\text{file}}$, based on the rank of the ground-truth buggy file in its prediction list.

    \item \textbf{Function-Level Reward ($r_{\text{func}}$):} The function-level agent is rewarded only if the file-level agent's prediction was correct. If this condition is met, it receives a reward, $r_{\text{func}}$, based on the rank of the correct buggy function.

    \item \textbf{Line-Level Reward ($r_{\text{line}}$):} The line-level agent is rewarded only if both the file and function predictions were correct. If this condition is met, it receives a reward, $r_{\text{line}}$, based on the rank of the correct buggy line.
\end{itemize}

The total reward, $R_{\text{total}}$, for a full localization sequence is a weighted sum of the rewards from each agent:
\[
R_{\text{total}} = \lambda_{\text{file}} \cdot r_{\text{file}} + \lambda_{\text{func}} \cdot r_{\text{func}} + \lambda_{\text{line}} \cdot r_{\text{line}}
\]

Here, $\lambda_{\text{file}}$, $\lambda_{\text{func}}$, and $\lambda_{\text{line}}$ are hyperparameters used to balance the contribution of each agent to the overall learning signal. For joint strategies, including VT and the joint fine-tuning stage of PT, we optimize the hierarchy using the composite reward $R_{\text{total}}$, with a shared critic estimating state values for advantage computation across levels. In contrast, under the decoupled strategies (TWF and T1B1), each agent is optimized independently via per-position REINFORCE using a local reward at its own granularity; TWF uses oracle parent inputs, whereas T1B1 uses frozen upstream predictions. Each decoupled agent uses its own critic conditioned only on the bug report embedding as an unbiased baseline.

\paragraph{Baseline Sparse Reward Strategy.}
For comparison, we also evaluate a traditional \textbf{sparse reward} strategy. This approach provides a non-zero reward only if the entire prediction chain—file, function, and line—is correct. The reward function is defined as:
\[
R_{\text{sparse}} = \begin{cases}
1 & \text{if file, function, and line predictions are all correct} \\
0 & \text{otherwise.}
\end{cases} \label{eq:sparse-reward}
\]
While simple, this strategy suffers from an infrequent and less informative feedback signal, which can hinder learning efficiency. Its performance is analyzed in our ablation studies in Section~\ref{subsubsec:reward-strategies}.

\subsection{Training Strategy}
\label{subsec:training}

Training a hierarchy of agents is notoriously challenging due to the problem of \textbf{compounding errors}: incorrect predictions from higher-level agents propagate down the hierarchy, providing noisy and misleading input for subsequent agents. This inherent instability can severely hinder or even prevent effective learning.

To overcome this, we experimented with multiple training protocols (as detailed in Section~\ref{subsubsec:training-strategies}) and found that a purely \textbf{Training with Teacher Forcing (TWF)} strategy provided strong and consistent results. This approach decouples the agents during the training phase, ensuring that each one learns from a clean, stable, and accurate signal. It entirely avoids the issue of compounding errors by never training a downstream agent on the noisy predictions of an upstream agent.

The training process is as follows:

\begin{enumerate}
    \item \textbf{File-Level Agent Training:} First, the file-level agent is trained independently using the complete dataset of bug reports. Its sole objective is to learn a policy that maps a bug report to the correct source file.

    \item \textbf{Function-Level Agent Training:} Next, the function-level agent is trained. Critically, during its training, it is \textbf{always} provided with the ground-truth buggy file, regardless of what the file-level agent would have predicted. This form of teacher forcing ensures the function-level agent can focus exclusively on learning to identify the correct function within a perfect context.

    \item \textbf{Line-Level Agent Training:} Finally, the line-level agent is trained. Similarly, it is always provided with both the ground-truth file and the ground-truth function. This allows it to learn the fine-grained task of line localization without being penalized by any potential upstream errors.
\end{enumerate}

By training each agent on a perfectly curated set of inputs, this sequential strategy ensures that every component of the hierarchy becomes a strong and specialized expert in its specific task. While the agents operate as a cascade during inference, this training method allows each agent to learn its task independently, contributing to the strong performance of our final model. Concretely, each agent is optimized via per-position REINFORCE with a per-agent critic baseline, using teacher-forced inputs to ensure each agent trains on clean, correct context.

\subsection{Implementation Details}
\label{subsec:hrl-impl}
All agents are trained using Adam with a learning rate of $1\times10^{-4}$ and a batch size of 32, for up to 40 epochs per phase. During training, agents use epsilon-greedy exploration with $\varepsilon$ annealed from 0.3 to 0.05 (decay factor 0.995 per step), reset at the start of each phase. The file-level agent ranks up to $N{=}5$ candidates; function- and line-level agents rank up to 15 candidates each. For intermediate rewards, the first five position weights are $[1.0, 0.8, 0.6, 0.4, 0.2]$, with rewards decreasing geometrically for lower-ranked positions. Incorrect candidates ranked above the correct target receive a small penalty of $-0.1$.

%% file: figures/architecture_diagram.tex
\begin{figure*}[t]
  \centering
  \includegraphics[width=\textwidth]{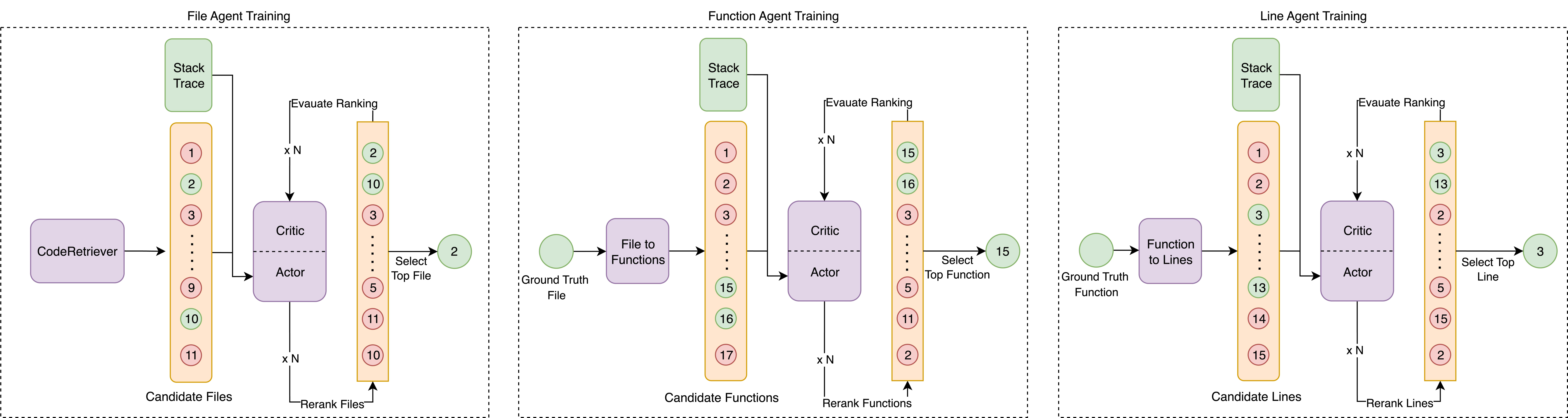}
  \caption{C2C hierarchical training pipeline. The diagram shows the sequential training strategy (TWF), where agents are trained using ground-truth oracle inputs to ensure stability.}
  \Description{A multi-stage pipeline diagram illustrating the C2C hierarchical training process. It begins with ground-truth/oracle inputs, trains agents in a teacher-forced or sequential (TWF) manner to maintain stability, and produces a final trained hierarchy of agents.}
  \label{fig:atlas_arch}
\end{figure*}

%% file: sections/results.tex
\section{Experiments and Results}
\label{sec:results}

We evaluated \texttt{C2C} against a suite of metrics and two baselines: (1) RLocator and (2) LLMAO. In the following section, we detail our experimental setup, selected evaluation metrics, and show the results of each methodology at the file, function, and line levels, where applicable. We also performed ablation studies to better understand the importance of each component in our proposed framework.

\subsection{Evaluation Setup}
\label{subsec:evaluation-setup}

\subsubsection{Baselines}
We evaluated \texttt{C2C} against two baselines representing different localization granularities. 

\paragraph{RLocator}
First, we compare against \textbf{RLocator} \cite{Chakraborty_2024_RL_RLocator}, an \gls{rl} method capable of bug localization at the file level, but which lacks finer-grained resolution. This baseline provides a strong benchmark for single-stage \gls{rl}-based localization. RLocator formulates bug localization as an \gls{mdp} and learns to rerank candidate files using CodeBERT representations and a CNN+LSTM ranking model trained with A2C, but its output remains a ranked list of files.

\paragraph{LLMAO}
Second, we compare against \textbf{LLMAO} \cite{Yang_2024_LLMAO}, a decoder-only language model-based approach designed explicitly for fine-grained, line-level bug localization. This comparison illustrates performance at the opposite end of the localization granularity spectrum. Given a buggy file, LLMAO predicts the buggy line over the entire file, skipping function level localization and performing strict line-level prediction. We therefore use LLMAO and determine its top buggy lines, aggregating across these predictions to derive function-level hit rates and MRR. LLMAO is built upon the 350M-parameter CodeGen model \cite{nijkamp2023codegen}, making it a relevant point of comparison for our CodeBERT-based retrieval and HRL localization framework. Because CodeGen is limited to 2,048 input tokens, LLMAO uses a 128-line sliding window with a random offset during training to control context length and reduce positional bias. As a result, LLMAO operates over a substantially smaller effective search space, considering the median lines per buggy file for Elasticsearch and SWE-bench Verified are 327 and 670, respectively.

\subsubsection{Evaluation Strategies and Metrics}
\label{subsec:evaluation}

We assess our model's performance using two distinct evaluation protocols, each designed to probe different aspects of our hierarchical framework. The final performance for each bug report, identified by a unique \textbf{issue ID}, is measured using a consistent set of ranking-based metrics.

\paragraph{Evaluation Protocols}
\begin{itemize}

    \item \textbf{End-to-End (E2E):} This protocol measures the end-to-end performance of the \texttt{C2C} hierarchical localization framework, with the ground-truth file present in the candidate set. The agents operate sequentially: the file-agent produces a ranked list of files and selects its top-ranked file, the function-agent ranks functions within the selected file and selects its top-ranked function, and the line-agent ranks lines within the selected function. This sequential evaluation captures error propagation across the hierarchy, as downstream agents operate on the predictions made by upstream agents. Line-level metrics are computed when the selected function contains an annotated buggy line. CodeRetriever's standalone retrieval performance is evaluated separately using its unmodified rankings.

    \item \textbf{Oracle:} In the ORACLE setting, downstream localization is evaluated under perfect upstream context. Concretely, the function-level agent is evaluated with the ground-truth buggy file, and the line-level agent is evaluated with the ground-truth buggy file and buggy functions, so errors from earlier stages do not propagate downward. This protocol isolates the ranking ability of each agent at its own granularity and measures performance assuming the higher-level decision was correct. Unlike the end-to-end protocol, ORACLE therefore reflects the upper bound of the hierarchical localizers under ideal parent selection.

    \item \textbf{LLMAO:} To facilitate a direct comparison to the LLMAO baseline, this protocol provides the \textbf{ground-truth file} directly to the function-level agent. For each issue ID, the evaluation then proceeds to rank the functions within that correct file, and subsequently the lines within those functions. The final rank is determined by the highest-ranking path (function and line) that matches the ground truth. This provides a focused assessment of the function and line-level agents' performance, independent of any file-level prediction errors.
\end{itemize}

\paragraph{Metrics}
The final rank generated by each protocol for an issue ID is used to compute the following standard performance metrics:
\begin{itemize}
    \item \textbf{Mean Reciprocal Rank (MRR):} The average reciprocal rank of the first correct prediction across the evaluated instances. This metric rewards models that place the correct bug location higher in the final ranked list.

    \item \textbf{Hits@K:} The percentage of evaluated bug-location instances for which a correct bug location is identified within the top-$K$ positions of the ranked list (for $K \in \{1,5,10\}$).
    
    \item \textbf{Recall@K:} Used specifically for the file-retrieval stage, this measures the percentage of unique issues for which at least one ground-truth buggy file is retrieved within the top-$K$ candidate files.
\end{itemize}

\subsection{Results}
\label{subsec:our-results}

\input{tables/accuracy_elasticsearch}
\input{tables/accuracy_swebench}

Table~\ref{tab:comparison} presents a comprehensive performance comparison of our \texttt{C2C} framework against the RLocator and LLMAO baselines on the Elasticsearch dataset. The results reported for \texttt{C2C} utilize our most effective sequential training strategy (TWF). All reported results are averaged across \textbf{3 random seeds}. It is important to preface in comparing baseline results a significant methodological limitation of LLMAO across both Elasticsearch and SWE-bench Verified evaluations. LLMAO's reliance on the CodeGen model family restricts its input context window to 2,048 tokens, effectively limiting its analysis to files no larger than 128 lines of code. Accordingly, incoming files provided to LLMAO have been truncated -- therefore LLMAO's reported performance may be inflated relative to C2C, which is evaluated on the complete file.

\paragraph{\textbf{Elasticsearch Results.}}
In the \textbf{End-to-End (E2E)} evaluation, our model demonstrates robust performance across the hierarchy. \texttt{C2C}, trained with TWF training strategy, achieves a file-level MRR of 0.5005 and correctly identifies the buggy file as its top prediction in 36.62\% of cases (Hit@1). Following this, it achieves a function-level MRR of 0.5450 and a Hit@1 of 43.66\%, and finally, a line-level MRR of 0.2932 with a Hit@1 of 6.5\%. For a detailed breakdown including other metrics like Hits@5 and Hits@10, please refer to Table~\ref{tab:comparison}.

To ensure a fair and direct comparison with the LLMAO baseline, we use the \textbf{LLMAO} evaluation, which isolates the performance of the downstream agents. Under these conditions, \texttt{C2C} (trained with TWF training) achieves a function-level MRR of 0.6574 (Hit@1 of 53.52\%) and a line-level MRR of 0.2251 (Hit@1 of 6.2\%). These results outperform the reported LLMAO baseline on function level metrics, scoring function-level Hit@1 of 4.90\% while at the line level, LLMAO achieves an MRR of 0.2149 compared with C2C's 0.2251. A full comparison can be found in Table \ref{tab:comparison}.

\paragraph{\textbf{SWE-bench Verified Results.}}
Table~\ref{tab:comparison_swebench} reports \texttt{C2C}'s performance on the SWE-bench Verified dataset, which spans 11 Python repositories. In the E2E evaluation, the TWF strategy achieves a file-level Hit@1 of 21.74\% and MRR of 0.4156, a function-level MRR of 0.2815, and a line-level MRR of 0.1000. The substantially lower file-level Hit@1 compared to Elasticsearch (36.62\%) may reflect the differences between Python and Java-based bug localization. Prior work notes that the dynamic nature of Python may affect the effectiveness of classic bug localization techniques \cite{Rezaalipour_Furia_2024}, especially stack trace-based methods, with Java stack traces often exposing rich identifiers such as package, class, and method names \cite{Java_stack_traces}. Nevertheless, the file-level Hit@5 of 76.09\% shows that the file agent ranks the correct file within its top five candidates in most cases. Under the LLMAO-style evaluation (ground-truth file provided), function Hit@1 improves to 23.91\% while line MRR drops to 0. Considering the median lines per buggy file for SWE-bench Verified is 670 (Table~\ref{tab:dataset-comparison}), LLMAO operates over a reduced context, resulting in a smaller effective search space. LLMAO achieves a function Hit@1 of 4.39\% and a line MRR of 0.1149.

\paragraph{\textbf{Computational Cost Analysis.}}
As C2C introduces a multi-stage retrieval and hierarchical localization pipeline, it is important to assess whether its added complexity imposes a practical latency burden. In bug localization, inference speed matters because the system is intended to support interactive developer workflows, where delays can reduce usability. Although our ablations (Section 7.3) show that the additional components of C2C provide measurable performance gains, these gains are only meaningful in practice if they do not come at prohibitive runtime cost. We therefore evaluate inference latency to determine whether C2C remains practical for real-world use.   

At inference time, the full \texttt{C2C} pipeline---file ranking, function extraction, and line ranking---processes a bug report in approximately 840ms on average (p50: 810\,ms, p95: 1.6\,s) on a single H100 GPU. The latency is dominated by CodeBERT embedding computation for candidate files and functions; the three-agent ranking cascade itself adds negligible overhead ($<$5\,ms), as the ranking heads are lightweight MLPs.

\input{tables/multi_file_line_coverage}
\paragraph{\textbf{Coverage of Multi-Location Bugs.}}

Many real-world bugs span multiple files or lines. Although \texttt{C2C} is trained to identify a single correct target at each hierarchical level, the ranked lists produced by each agent provide natural coverage over multiple locations. We measure \textbf{multi-file recall} at two stages: the percentage of issues for which at least one ground-truth buggy file is retrieved within the top-10 by CodeRetriever (results in Table~\ref{tab:retrieval_metrics}), and the fraction ranked within the top-5 by the file agent. We further measure \textbf{multi-line recall} as the fraction of all ground-truth buggy lines recovered by the full cascade (5 files $\times$ 15 functions $\times$ 15 lines per issue). Results are shown in Table~\ref{tab:multi_location}.

The file-agent evaluation yields a top-5 coverage of 54.29\% on Elasticsearch and 66.25\% on SWE-bench Verified, compared with CodeRetriever's top-10 multi-file recall of 36.36\% and 65.00\%, respectively. Since these results are computed at different stages of the pipeline and under different candidate-set representations, they should not be interpreted as a direct stage-by-stage improvement in retrieval recall. Multi-line recall is 17.58\% and 12.50\%, respectively, reflecting the increasing selectivity of the three-stage pipeline where only lines within highly ranked functions of highly ranked files are considered.

We emphasize that this multi-location coverage is \emph{emergent} from the top-$K$ ranking mechanism and is not an explicit training objective --- each agent is optimized to rank a single correct target highly, and any additional ground-truth locations that receive high scores are a byproduct of the learned representations. Extending \texttt{C2C} to explicitly optimize for multi-target recall, for example through set-based reward functions or multi-label ranking losses, is a promising direction for future work.

\subsection{Ablation Studies}
\label{subsec:ablation}
To better quantify the contributions of individual components within \texttt{C2C}, we conducted a series of ablation studies. These experiments systematically isolate key architectural and training elements to assess their effect on overall system performance.

\input{tables/iterative_contrastive_loss_ablation}
\subsubsection{Impact of Contrastive Fine-Tuning and Hard Negative Mining}
We first evaluate the effect of iterative contrastive loss fine-tuning and hard negative mining on CodeRetriever's performance. As shown in Table~\ref{tab:retrieval_metrics}, without contrastive fine-tuning, both the hit rate and recall are negligible on both datasets. After the first training iteration with triplet loss, performance improves substantially.

On the \textbf{Elasticsearch} dataset (Panel~A), the second iteration, which incorporates \textbf{hard negative mining} using a \gls{faiss} index, further boosts retrieval quality over the initial model, improving \textbf{Hit@10} from 55.93\% to 57.84\% and \textbf{Recall@10} from 33.33\% to 36.36\%, eventually reaching a Hit@50 of 72.88\%.

On the \textbf{SWE-bench Verified} dataset (Panel~B), the iterative procedure yields substantial gains at the K values critical for downstream HRL: hard-negative mining improves Hit@5 from 57.24\% to 60.00\% and Hit@10 from 62.07\% to 69.66\%. Absolute retrieval performance is higher on SWE-bench Verified than on Elasticsearch at these depths (e.g., Hit@10 of 69.66\% vs. 57.84\%), despite the multi-repository setting. Overall, the progression from the base model to the hard-negative model is consistent across both datasets, confirming the value of iterative contrastive training.

\subsubsection{Effect of Hierarchical Structure and Encoders on Localization Accuracy}

We investigate the role of key architectural components in \texttt{C2C}'s performance, as shown in Table~\ref{tab:network_ablation}. Removing the LSTM encoder, which models sequential dependencies among candidate embeddings, degrades function- and line-level localization across both datasets: on Elasticsearch, function-level \gls{mrr} drops from 0.5450 to 0.4893, and on SWE-bench Verified the line-level \gls{mrr} collapses from 0.1000 to 0.0000. The mixed effect at the file level (a slight gain on Elasticsearch, a drop on SWE-bench Verified) suggests that file-level ranking is less reliant on sequential context than finer-grained localization, where ordering among candidates carries a stronger inductive signal.

We further examine the effect of the hierarchy itself by comparing \texttt{C2C} against a flat variant that predicts lines directly, bypassing the file-and-function decomposition. This ablation produces a dramatic drop in line-level \gls{mrr} on both datasets---from 0.2932 to 0.0682 on Elasticsearch and from 0.1000 to 0.0000 on SWE-bench Verified---demonstrating the importance of the hierarchical structure for fine-grained localization. Together, these results establish that both the contextual LSTM encoding and the hierarchical decomposition are important components of C2C's architecture.
\input{tables/atlas_network_components}

\subsubsection{Impact of Reward Shaping: Intermediate vs.\ Sparse Rewards}
\label{subsubsec:reward-strategies}

We assess the impact of level-specific feedback on hierarchical learning by comparing \textbf{intermediate} and \textbf{sparse} reward strategies across both datasets, as shown in Table~\ref{tab:reward_ablation}. On Elasticsearch, intermediate rewards consistently outperform sparse rewards at every level of the hierarchy: file \gls{mrr} improves from 0.5005 to 0.5225, function \gls{mrr} from 0.5223 to 0.5450, and line \gls{mrr} from 0.2377 to 0.2932.

The pattern is even more pronounced on SWE-bench Verified: while intermediate rewards yield a line-level \gls{mrr} of 0.1000, the sparse strategy fails entirely at the line level, achieving an \gls{mrr} of 0.0000. This finding is consistent across both Java and Python settings and highlights a fundamental challenge of sparse reward \gls{rl} in hierarchical localization: credit assignment across three sequential agents becomes particularly challenging when the reward signal is withheld until the full prediction chain is correct. Intermediate rewards resolve this by providing a stable, informative signal at each stage of the hierarchy, allowing agents to learn effective policies independently before the joint cascade is evaluated.

\subsubsection{Comparison of Training Strategies for Hierarchical Agents}
\label{subsubsec:training-strategies}
The strategy used to train a hierarchical model is critical, as the sequential dependency between agents can lead to compounding errors. A poorly trained upstream agent provides a noisy, non-stationary learning signal to downstream agents, which can severely destabilize and hinder the learning process. We systematically evaluated four distinct training strategies:
\input{tables/atlas_reward_mechanism}
\begin{itemize}
    \item \textbf{Vanilla Training (VT):} A naive end-to-end approach where all three agents are trained jointly from random initialization. The output of the file agent is passed directly to the function agent, and its output to the line agent, with gradients flowing back through the entire hierarchy. This strategy performs poorly because the function and line agents are initially fed incoherent, near-random inputs, preventing them from learning a meaningful policy.

    \item \textbf{Pretraining Agents Sequentially (PT):} This strategy mitigates the instability of VT by first pretraining each agent sequentially and then fine-tuning them jointly. While pretraining provides a better starting point, the joint fine-tuning phase re-introduces instability: gradients from the still-learning downstream agents can disrupt the already well-trained upstream agents, leading to suboptimal results.

    \item \textbf{Training One-by-One (T1B1):} The file-level agent is trained until convergence and its weights are then frozen. The predictions from this frozen agent serve as training data for the function-level agent, which is subsequently frozen, and so on, with each agent optimized via per-position REINFORCE. This ensures a stable learning signal for downstream agents, as the upstream policies do not change during their training.

    \item \textbf{Training with Teacher Forcing (TWF):} Each agent is trained independently using ground-truth oracle selections from the preceding levels, optimized via per-position REINFORCE. The function-level agent is always trained using the correct ground-truth file, and the line-level agent is always trained using the correct ground-truth file and function. This completely decouples the agents during training, guaranteeing that every agent learns from a perfect and stable signal.
\end{itemize}

\input{tables/atlas_training_strategy}

To isolate each agent's ranking ability from upstream errors, we evaluate all strategies under an oracle protocol where the ground-truth candidate is always present in the candidate pool. This directly measures what each agent has learned assuming the upper-level agent made the correct decision. Since all strategies operate over the same candidate set at the file level, file-level metrics are identical across all strategies by construction, and the strategies only diverge meaningfully at the function and line levels. Table ~\ref{tab:training_strategy} presents the evaluation results using the four training strategies across both datasets.

From the table, we observe that at the \textbf{function level} T1B1 achieves the strongest performance on Elasticsearch (Hit@1 0.5500, MRR 0.6376), while TWF dominates on SWE-bench Verified across all metrics (Hit@1 0.3214, MRR 0.4286), substantially ahead of T1B1 (MRR 0.3817). This suggests that while freezing upstream agents provides a stable context on a single repository, training against perfect oracle inputs is more effective in the harder multi-repository setting where greater variability in code structure and bug patterns demands a cleaner learning signal.

At the \textbf{line level}, no single strategy dominates across both datasets. On \textbf{Elasticsearch}, VT attains the highest MRR (0.4027) and Hit@10 (0.7643), while TWF achieves the best Hit@1 (0.2323). On \textbf{SWE-bench Verified}, T1B1 obtains the highest MRR (0.2897), while TWF achieves the best Hit@10 (0.6207). Despite these mixed line-level results, TWF and T1B1 remain the most competitive strategies overall, and they do so with a much cleaner and more stable training signal than VT or PT.

Because TWF and T1B1 emerge as the two strongest decoupled strategies, we compare them directly using paired statistical significance testing. Specifically, we apply the two-sided Wilcoxon signed-rank test~\cite{wilcoxon1945} with $\alpha = 0.05$ to reciprocal rank (RR) scores under the oracle evaluation protocol. Scores are pooled across three random seeds (0, 20, 42), yielding $n=213$ paired observations for Elasticsearch (71 instances $\times$ 3 seeds) and $n=138$ for SWE-bench Verified (46 instances $\times$ 3 seeds). File-level performance is not tested, since TWF and T1B1 train the file-level agent identically and therefore produce identical scores by construction.

\input{tables/statistical_testing_twf}

Table~\ref{tab:twf_t1b1_significance} shows that, on \textbf{Elasticsearch} at the \textbf{function level}, T1B1 attains a slightly higher mean RR than TWF (0.7215 vs.\ 0.7047), but the difference is not statistically significant ($p = 0.507$; $\delta = -0.009$, negligible). At the \textbf{line level}, TWF significantly outperforms T1B1 (0.6100 vs.\ 0.5432; $p = 0.002$), although the effect size remains negligible ($\delta = 0.141$). On \textbf{SWE-bench Verified}, TWF achieves higher mean RR than T1B1 at both the \textbf{function level} (0.4099 vs.\ 0.3903) and the \textbf{line level} (0.2930 vs.\ 0.2771), but neither difference is statistically significant ($p = 0.173$ and $p = 0.361$, respectively). Overall, the effect sizes are negligible across all comparisons, indicating that the absolute performance gaps are small, but the direction of the differences generally favors TWF.

\noindent\textit{Note.} The mean RR values reported in Table~\ref{tab:twf_t1b1_significance} may differ slightly from those in Table~\ref{tab:training_strategy}. The training-strategy table reports micro-averaged MRR over evaluations, whereas the statistical test uses instance-level RR scores pooled across the three random seeds, with each HRL instance contributing one paired observation per seed. When an instance contains multiple valid localization paths, we use the best RR across those paths.

Taken together, these results support using \textbf{TWF} as the primary training strategy for \textsc{C2C}. Although the differences between TWF and T1B1 are usually small, TWF is more consistent across datasets, achieves the strongest function-level performance on SWE-bench Verified, and yields the clearest statistically significant gain at the Elasticsearch line level.

%% file: tables/accuracy_elasticsearch.tex
\begin{table*}[t]
\centering
\caption{Comparison of \texttt{C2C} with baseline methods for localization accuracy (\%) on the Elasticsearch dataset (single Java repository). LLMAO-style evaluation provides the ground-truth file to the agent. E2E results reflect sequential end-to-end performance across the hierarchical localization agents, with the ground-truth file present in the candidate set. Results are averaged across 3 random seeds.}
\label{tab:comparison}
\resizebox{\textwidth}{!}{
\begin{tabular}{lccccccc}
\toprule
\textbf{Method} & \textbf{File Hit@1} & \textbf{File Hit@5} & \textbf{File MRR} & \textbf{Func Hit@1} & \textbf{Func MRR} & \textbf{Line Hit@10} & \textbf{Line MRR} \\ \midrule
RLocator & \textbf{36.62} & -- & -- & -- & -- & -- & -- \\
LLMAO & -- & -- & -- & 4.90 & 0.0617 & 41.76 & 0.2149 \\
\textbf{\texttt{C2C} (TWF, E2E)} & \textbf{36.62} & \textbf{80.28} & \textbf{0.5005} & 43.66 & 0.5450 & 54.55 & 0.2932 \\
\textbf{\texttt{C2C} (TWF, ORACLE)} & \textbf{36.62} & \textbf{80.28} & \textbf{0.5005} & \textbf{54.55} & 0.6298 & \textbf{73.06} & \textbf{0.3863} \\
\textbf{\texttt{C2C} (TWF, LLMAO Style)} & -- & -- & -- & 53.52 & \textbf{0.6574} & 58.33 & 0.2251 \\
\bottomrule
\end{tabular}
}
\end{table*}

%% file: tables/accuracy_swebench.tex
\begin{table*}[t]
\centering
\caption{Comparison of \texttt{C2C} with baseline methods for localization accuracy (\%) on the SWE-bench Verified dataset (11 Python repositories). LLMAO-style evaluation provides the ground-truth file to the agent. E2E results reflect sequential end-to-end performance across the hierarchical localization agents, with the ground-truth file present in the candidate set. Results are averaged across 3 random seeds.}
\label{tab:comparison_swebench}
\resizebox{\textwidth}{!}{
\begin{tabular}{lccccccc}
\toprule
\textbf{Method} & \textbf{File Hit@1} & \textbf{File Hit@5} & \textbf{File MRR} & \textbf{Func Hit@1} & \textbf{Func MRR} & \textbf{Line Hit@10} & \textbf{Line MRR} \\ \midrule
RLocator & \textbf{21.74} & -- & -- & -- & -- & -- & -- \\
LLMAO & -- & -- & -- & 4.39 & 0.0719 & 23.03 & 0.1149 \\
\textbf{\texttt{C2C} (TWF, E2E)} & \textbf{21.74} & \textbf{76.09} & \textbf{0.4156} & 19.57 & 0.2815 & 50.00 & 0.1000 \\
\textbf{\texttt{C2C} (TWF, ORACLE)} & \textbf{21.74} & \textbf{76.09} & \textbf{0.4156} & \textbf{32.14} & \textbf{0.4298} & \textbf{51.72} & \textbf{0.2846} \\
\textbf{\texttt{C2C} (TWF, LLMAO Style)} & -- & -- & -- & 23.91 & 0.3548 & 0 & 0 \\
\bottomrule
\end{tabular}
}
\end{table*}

%% file: tables/multi_file_line_coverage.tex
\begin{table}[t]
\centering
\caption{Multi-location coverage of the TWF-trained \texttt{C2C} model. The Retrieval column reports the percentage of unique issues for which at least one ground-truth buggy file is retrieved within the top-10 by CodeRetriever. The File Agent column reports top-5 coverage under the HRL candidate-set evaluation. Multi-line recall measures the fraction of ground-truth buggy lines recovered by the full cascade (5 files $\times$ 15 functions $\times$ 15 lines).}

\label{tab:multi_location}
\begin{tabular}{lcccc}
\toprule
& \multicolumn{2}{c}{\textbf{Multi-File Recall}} & \multicolumn{1}{c}{\textbf{Multi-Line Recall}} \\
\cmidrule(lr){2-3} \cmidrule(lr){4-4}
\textbf{Dataset} & \textbf{Retrieval} & \textbf{File Agent} & \textbf{Full Cascade} \\
\midrule
Elasticsearch   & 36.36 & 54.29 & 17.58 \\
SWE-bench       & 65.00 & 66.25 & 12.50 \\
\bottomrule
\end{tabular}
\end{table}

%% file: tables/iterative_contrastive_loss_ablation.tex
\begin{table*}[t]
  \caption{Impact of iterative contrastive fine-tuning on retrieval performance (test set). Hit@K: percentage of evaluated bug-location instances for which at least one ground-truth buggy file associated with the issue appears in the top-$K$. Recall@K: percentage of unique issues for which at least one ground-truth buggy file is retrieved within the top-$K$. Panel~A shows the Elasticsearch (Java) dataset; Panel~B shows SWE-bench Verified (Python, 11 repos).}
  \label{tab:retrieval_metrics}
  \centering
  \small

  %% Panel A: Elasticsearch
  \textbf{Panel A: Elasticsearch}
  \vspace{0.3em}

  \resizebox{\textwidth}{!}{%
  \begin{tabular}{l*{5}{cc}}
    \toprule
    \textbf{Training configuration}
      & \multicolumn{2}{c}{\textbf{@3}}
      & \multicolumn{2}{c}{\textbf{@5}}
      & \multicolumn{2}{c}{\textbf{@10}}
      & \multicolumn{2}{c}{\textbf{@20}}
      & \multicolumn{2}{c}{\textbf{@50}} \\
    & \textbf{Hit} & \textbf{Recall}
    & \textbf{Hit} & \textbf{Recall}
    & \textbf{Hit} & \textbf{Recall}
    & \textbf{Hit} & \textbf{Recall}
    & \textbf{Hit} & \textbf{Recall} \\
    \midrule
    No contrastive loss            &  0.16 &  0.04 &  0.16 &  0.04 &  0.16 &  0.04 &  3.29 &  0.83 &  6.18 &  1.56 \\
    1st-iteration contrastive loss & 23.94 & \textbf{15.15} & \textbf{41.95} & \textbf{21.21} & 55.93 & 33.33 & 56.57 & 36.36 & \textbf{75.21} & \textbf{57.58} \\
    2nd-iteration contrastive loss & \textbf{29.66} & \textbf{15.15} & 41.53 & \textbf{21.21} & \textbf{57.84} & \textbf{36.36} & \textbf{61.23} & \textbf{39.39} & 72.88 & 51.52 \\
    \bottomrule
  \end{tabular}%
  }

  \vspace{1em}

  %% Panel B: SWE-bench Verified
  \textbf{Panel B: SWE-bench Verified}
  \vspace{0.3em}

  \resizebox{\textwidth}{!}{%
  \begin{tabular}{l*{5}{cc}}
    \toprule
    \textbf{Training configuration}
      & \multicolumn{2}{c}{\textbf{@3}}
      & \multicolumn{2}{c}{\textbf{@5}}
      & \multicolumn{2}{c}{\textbf{@10}}
      & \multicolumn{2}{c}{\textbf{@20}}
      & \multicolumn{2}{c}{\textbf{@50}} \\
    & \textbf{Hit} & \textbf{Recall}
    & \textbf{Hit} & \textbf{Recall}
    & \textbf{Hit} & \textbf{Recall}
    & \textbf{Hit} & \textbf{Recall}
    & \textbf{Hit} & \textbf{Recall} \\
    \midrule
    No contrastive loss            &  0.00 &  0.00 &  0.00 &  0.00 &  0.00 &  0.00 &  0.00 &  0.00 &  0.69 &  2.50 \\
    1st-iteration contrastive loss & \textbf{52.41} & \textbf{35.00} & 57.24 & 45.00 & 62.07 & 55.00 & 80.00 & 67.50 & 91.03 & 77.50 \\
    2nd-iteration contrastive loss & 50.34 & \textbf{35.00} & \textbf{60.00} & \textbf{47.50} & \textbf{69.66} & \textbf{65.00} & \textbf{87.59} & \textbf{77.50} & \textbf{93.79} & \textbf{85.00} \\
    \bottomrule
  \end{tabular}%
  }
\end{table*}

%% file: tables/atlas_network_components.tex
\begin{table}[t]
\centering
\caption{Impact of Network Components on \texttt{C2C} Performance (MRR).}
\label{tab:network_ablation}
\begin{tabular}{@{}llcccccc@{}}
\toprule
& & \multicolumn{3}{c}{\textbf{Elasticsearch}} & \multicolumn{3}{c}{\textbf{SWE-bench Verified}} \\
\cmidrule(lr){3-5} \cmidrule(lr){6-8}
\textbf{Component} & \textbf{Metric} & \textbf{Base} & \textbf{Ablated} & \textbf{Change} & \textbf{Base} & \textbf{Ablated} & \textbf{Change} \\
\midrule
\multirow{3}{*}{LSTM Encoder} & File MRR & 0.5225 & 0.5653 & +0.0428 & 0.4112 & 0.3627 & $-$0.0485 \\
                               & Func MRR & 0.5450 & 0.4893 & $-$0.0557 & 0.2259 & 0.2165 & $-$0.0094 \\
                               & Line MRR & 0.2932 & 0.2925 & $-$0.0007 & 0.1000 & 0.0000 & $-$0.1000 \\
\midrule
Hierarchy                      & Line MRR & 0.2932 & 0.0682 & $-$0.2250 & 0.1000 & 0.0000 & $-$0.1000 \\
\bottomrule
\end{tabular}
\end{table}

%% file: tables/atlas_reward_mechanism.tex
\begin{table}[t]
\centering
\caption{Effect of Reward Mechanism on \texttt{C2C} Performance (MRR).}
\label{tab:reward_ablation}
\begin{tabular}{@{}lcccc@{}}
\toprule
& \multicolumn{2}{c}{\textbf{Elasticsearch}} & \multicolumn{2}{c}{\textbf{SWE-bench Verified}} \\
\cmidrule(lr){2-3} \cmidrule(lr){4-5}
\textbf{Metric} & \textbf{Intermediate} & \textbf{Sparse} & \textbf{Intermediate} & \textbf{Sparse} \\ \midrule
File MRR     & \textbf{0.5225} & 0.5005 & \textbf{0.4112} & 0.3413 \\
Function MRR & \textbf{0.5450} & 0.5223 & \textbf{0.2259} & 0.2122 \\
Line MRR     & \textbf{0.2932} & 0.2377 & \textbf{0.1000} & 0.0000 \\ \bottomrule
\end{tabular}
\end{table}

%% file: tables/atlas_training_strategy.tex
\begin{table*}[t]
\centering
\small
\caption{Comparison of training strategies using oracle evaluation (ground-truth candidate always
present). Elasticsearch results are from a single seed; SWE-bench results are averaged across 2
seeds. Best results per row and dataset are bolded.}
\label{tab:training_strategy}
\begin{tabular}{llcccccccc}
\toprule
& & \multicolumn{4}{c}{\textbf{Elasticsearch}} & \multicolumn{4}{c}{\textbf{SWE-bench Verified}} \\
\cmidrule(lr){3-6} \cmidrule(lr){7-10}
\textbf{Level} & \textbf{Metric} & \textbf{PT} & \textbf{VT} & \textbf{T1B1} & \textbf{TWF}
                                 & \textbf{PT} & \textbf{VT} & \textbf{T1B1} & \textbf{TWF} \\
\midrule
\multirow{4}{*}{File}
    & Hit@1  & 0.3662 & 0.3662 & 0.3662 & 0.3662 & 0.2174 & 0.2174 & 0.2174 & 0.2174 \\
    & Hit@5  & 0.8028 & 0.8028 & 0.8028 & 0.8028 & 0.7391 & 0.7391 & 0.7391 & 0.7391 \\
    & Hit@10 & 0.8028 & 0.8028 & 0.8028 & 0.8028 & 0.7391 & 0.7391 & 0.7391 & 0.7391 \\
    & MRR    & 0.5225 & 0.5225 & 0.5225 & 0.5225 & 0.4112 & 0.4112 & 0.4103 & 0.4112 \\
\cmidrule(l){2-10}
\multirow{4}{*}{Function}
    & Hit@1  & 0.3727 & 0.4091 & \textbf{0.5500} & 0.5455
             & 0.1339 & 0.1429 & 0.2500          & \textbf{0.3214} \\
    & Hit@5  & 0.7182 & 0.7227 & 0.7500          & \textbf{0.7636}
             & 0.3750 & 0.3482 & 0.5536          & \textbf{0.5714} \\
    & Hit@10 & 0.7909 & 0.7909 & \textbf{0.8136} & 0.8091
             & 0.6250 & 0.5536 & 0.6875          & \textbf{0.7321} \\
    & MRR    & 0.5247 & 0.5356 & \textbf{0.6376} & 0.6298
             & 0.2611 & 0.2483 & 0.3817          & \textbf{0.4286} \\
\cmidrule(l){2-10}
\multirow{4}{*}{Line}
    & Hit@1  & 0.2256 & 0.2290          & 0.1987 & \textbf{0.2323}
             & 0.1207 & \textbf{0.1983} & 0.1810 & 0.1638 \\
    & Hit@5  & \textbf{0.6263} & 0.5960 & 0.5522 & 0.5791
             & \textbf{0.4311} & 0.3621          & 0.4138 & 0.3707 \\
    & Hit@10 & 0.7508 & \textbf{0.7643} & 0.6869 & 0.7306
             & 0.5862 & 0.5604          & 0.5690 & \textbf{0.6207} \\
    & MRR    & 0.4004 & \textbf{0.4027} & 0.3530 & 0.3863
             & 0.2675 & 0.2793          & \textbf{0.2897} & 0.2773 \\
\bottomrule
\end{tabular}
\end{table*}

%% file: tables/statistical_testing_twf.tex
\begin{table}[t]
\centering
\caption{Paired statistical comparison between TWF and T1B1 using the Wilcoxon signed-rank test on per-instance reciprocal rank scores. Statistically significant results are shown in bold.}
\label{tab:twf_t1b1_significance}
\small
\begin{tabular}{llcccc}
\toprule
\textbf{Dataset} & \textbf{Level} & \textbf{TWF} & \textbf{T1B1} & \textbf{Cliff's $\delta$} & \textbf{$p$-value} \\
\midrule
Elasticsearch & Function & 0.7047 & 0.7215 & $-0.009$ & 0.507 \\
Elasticsearch & Line & 0.6100 & 0.5432 & $0.141$ & \textbf{0.002} \\
SWE-bench Verified & Function & 0.4099 & 0.3903 & $0.058$ & 0.173 \\
SWE-bench Verified & Line & 0.2930 & 0.2771 & $0.109$ & 0.361 \\
\bottomrule
\end{tabular}
\end{table}

%% file: sections/discussion.tex
\section{Discussion}
\label{sec:discussion}
Our research was guided by four key questions focused on evaluating the retrieval quality, hierarchical reasoning, and reward design of \texttt{C2C}. We address each question below, presenting our key findings and referencing the empirical evidence detailed in Section~\ref{sec:results}.

\subsection{RQ1: How effectively does CodeRetriever align stack traces and buggy files in the embedding space?}
\label{subsec:rq1}

\textbf{Key Findings:} Task-specific fine-tuning is essential for aligning code and natural language representations for bug localization. Our ablation studies (Section~\ref{subsec:ablation}) show that while a general-purpose pretrained model like CodeBERT is a good starting point, its initial alignment is weak. Iterative contrastive fine-tuning, enhanced with hard negative mining, proved critical to reshaping the embedding space. This process enables CodeRetriever to effectively distinguish between correct and incorrect code, thereby producing a compact, high-quality candidate set of files for the downstream localization agents.

\subsection{RQ2: How does hierarchical decomposition improve localization over single-stage methods?}
\label{subsec:rq2}

\textbf{Key Findings:} Hierarchical decomposition makes the complex problem of bug localization tractable. Our experiments demonstrate that hierarchical decomposition provides a substantial benefit for high-resolution localization. As shown in our ablation studies (Section~\ref{subsec:ablation}), removing the hierarchy and attempting to predict lines directly from the bug report causes a large drop in performance. By progressively narrowing the search space from files to functions and finally to lines, \texttt{C2C} improves fine-grained localization compared with the evaluated flat variant, ultimately providing developers with a more actionable starting point for debugging.

\subsection{RQ3: How do different reward strategies affect learning in hierarchical RL?}
\label{subsec:rq3}

\textbf{Key Findings:} Decomposed, level-specific feedback is important for training the hierarchical agents effectively. Our experiments comparing reward strategies (Section~\ref{subsubsec:reward-strategies}) show that providing \textbf{intermediate rewards} at each stage of the hierarchy consistently and substantially outperforms a single \textbf{sparse reward} given only for a correct final prediction. The sparse reward signal creates a difficult credit assignment problem that hinders learning. In contrast, intermediate rewards provide a stable, informative signal that guides each agent effectively---and on SWE-bench prove essential for any line-level learning at all, where sparse reward yields an MRR of 0.0 while intermediate rewards enable non-zero localization.

\subsection{RQ4: Does \texttt{C2C} generalize across programming languages and repositories?}
\label{subsec:rq4}

\textbf{Key Findings:} \texttt{C2C}'s architecture and training methodology generalize to a multi-repository Python setting, though absolute performance varies with task difficulty. Our evaluation on SWE-bench Verified---spanning 11 Python repositories---demonstrates that the key findings from the Elasticsearch (Java) evaluation hold: (1) iterative contrastive fine-tuning with hard negative mining consistently improves retrieval quality (Table~\ref{tab:retrieval_metrics}, Panel~B); (2) decoupled training strategies (TWF, T1B1) outperform joint strategies (VT, PT) across both datasets (Table~\ref{tab:training_strategy}); and (3) the hierarchical decomposition successfully localizes bugs at multiple granularity levels.

The file-level Hit@1 drops from 36.62\% (Elasticsearch) to 21.74\% (SWE-bench), which we attribute to three factors: (a)~the SWE-bench FAISS index covers \emph{all} files in each repository (thousands of candidates per query), whereas the Elasticsearch index contains only 6{,}017 files total; (b)~11 diverse Python repositories present greater variability in code structure, naming conventions, and bug patterns than a single Java project; and (c)~Python stack traces may carry less structural information (e.g., fewer package-qualified class names) than Java stack traces. Despite these challenges, the file-level Hit@5 of 76.09\% indicates that the correct file is usually among the top candidates, and the downstream agents remain effective when the file-level retrieval succeeds (LLMAO-style function Hit@1 of 23.91\%).

%% file: sections/threats_to_validity.tex
\section{Threats to Validity}
\label{sec:threats}

\subsection{Internal Validity}
This relates to the strength of the causal relationship between our method and the observed results. While we build on the original CodeBERT model trained in early 2020, we cannot rule out the possibility that portions of the Elasticsearch or SWE-bench source code, specifically code segments preceding the fix commits, were included in its pretraining data. Although the BEETLEBOX dataset was constructed in 2024 with safeguards to minimize data leakage, and the model predates it, this potential overlap remains a threat. To mitigate this, we encourage future work to further investigate and control for such leakage.

To prevent data leakage between training and evaluation, all dataset splits are performed at the \textbf{issue level} with stratification by repository, ensuring that all triplets derived from the same issue appear in the same partition and no stack trace information is shared across splits.

\subsection{External Validity}
This pertains to the generalizability of our findings. Our experiments span two complementary settings: a single large-scale Java project (Elasticsearch, 326 bug reports) and 11 Python repositories from SWE-bench Verified (393 issues). While this provides evidence of cross-language and cross-project generalizability, the evaluation remains limited to Java and Python. Although the training data includes multi-file, multi-function, and multi-line bugs, we assume during inference that each bug is localized to a single file, function, and line. This simplifying assumption may limit applicability in scenarios involving cross-file dependencies or distributed bugs. Future work should evaluate additional languages and bug types to further assess generalization. Finally, our evaluation uses CodeBERT as the retrieval encoder, and we do not assess alternative pretrained encoding models. As a result, it remains unclear whether our findings generalize to other backbone models, which may differ in their representation quality, cross-modal alignment ability, and retrieval effectiveness.

Finally, we selected TWF as the primary training strategy based on the empirical results observed on Elasticsearch and SWE-bench Verified. Although TWF showed the strongest and most consistent overall performance in our study, this choice may not generalize to other datasets, repositories, programming languages, or model backbones, where alternative strategies such as T1B1 could be preferable. Therefore, our conclusions regarding the superiority of TWF should be interpreted within the scope of the evaluated settings.

\subsection{Construct Validity}
This threat concerns whether our experimental setup and metrics faithfully capture the bug localization ability we intend to measure. Although we evaluate using line-level accuracy and \gls{mrr}, these metrics remain imperfect proxies for real developer utility, since they do not directly reflect debugging effort, interpretability, or how easily a ranked result can be acted upon in practice. In addition, our training objective uses position-based rewards over Top-$K$ rankings at the file, function, and line levels. While this better aligns learning with ranking quality than a purely binary reward design, it still provides a simplified approximation of bug localization usefulness. In particular, the reward structure imposes discrete rank thresholds and assigns no credit to candidates outside the Top-$K$, which may underrepresent partially useful predictions that are semantically close to the ground truth. Future work could explore more fine-grained and context-sensitive reward signals, such as rewards informed by call graph distance, shared dependencies, control/data flow proximity, or execution path similarity. Such extensions may better reflect the practical value of localization results, but would require additional program analysis and execution modeling beyond the scope of this work. Our E2E evaluation guarantees that the ground-truth file is present in the HRL file candidate set. Consequently, these results evaluate error propagation through the hierarchical localization agents but do not include failures caused solely by CodeRetriever omitting the correct file.

%% file: sections/conclusion.tex
\section{Conclusion}
\label{sec:conclusion}

This paper presents \texttt{C2C}, a novel hierarchical reinforcement learning (HRL) framework that localizes bugs at the file, function, and line-level. By combining a recall-focused retriever with a precision-focused hierarchical policy, \texttt{C2C} outperforms the compared baselines at the file level while extending localization to the function and line, offering more fine-grained debugging guidance.

Our ablation studies highlight the contribution of each component. The hierarchical structure improves line-level MRR by 4.3$\times$ on Elasticsearch and enables non-zero line localization on SWE-bench where the evaluated flat variant achieves an MRR of 0. Likewise, contrastive learning substantially improves retrieval performance, while intermediate, level-specific rewards consistently outperform sparse rewards across the evaluated settings.

While promising, \texttt{C2C} can be improved. Future work will focus on more granular reward mechanisms and more comprehensive code indexing strategies. \texttt{C2C} provides a step toward bridging semantic code understanding with strategic bug localization, and future work will extend these capabilities to cross-repository learning and integration with automated program repair.

\section*{Acknowledgments}

We acknowledge the support and funding of CIFAR, the Natural Sciences and Engineering Research Council of Canada (NSERC), and the Digital Research Alliance of Canada. This research was enabled through computational resources provided by the Research Computing Services group at the University of Calgary. We also thank Bilal Ahmed and Ali Mohammadi Ruzbahani for their initial work on preparing the Elasticsearch data from BEETLEBOX, including repository-level data extraction, parsing, initial filtering of data points, and extraction of stack-trace information.

%% file: references.bib
@article{ZouJavaFLSurvey,
  title={An empirical study of fault localization families and their combinations},
  author={Zou, Daming and Liang, Jingjing and Xiong, Yingfei and Ernst, Michael D and Zhang, Lu},
  journal={IEEE Transactions on Software Engineering},
  volume={47},
  number={2},
  pages={332--347},
  year={2019},
  publisher={IEEE}
}

@article{Rezaalipour_Furia_2024,
  title={An empirical study of fault localization in Python programs},
  author={Rezaalipour, Mohammad and Furia, Carlo A},
  journal={Empirical Software Engineering},
  volume={29},
  number={4},
  pages={92},
  year={2024},
  publisher={Springer}
}

@techreport{krasner2022cpsq,
  author       = {Herb Krasner},
  title        = {The Cost of Poor Software Quality in the U.S.: A 2022 Report},
  institution  = {Consortium for Information and Software Quality (CISQ)},
  year         = {2022},
  month        = nov,
  url          = {https://www.it-cisq.org/wp-content/uploads/sites/6/2022/11/CPSQ-Report-Nov-22-2.pdf},
  note         = {Accessed: 2026-03-07}
}

@inproceedings{BohmeFSE2017,
  title={Where is the bug and how is it fixed? an experiment with practitioners},
  author={B{\"o}hme, Marcel and Soremekun, Ezekiel O and Chattopadhyay, Sudipta and Ugherughe, Emamurho and Zeller, Andreas},
  booktitle={Proceedings of the 2017 11th joint meeting on foundations of software engineering},
  pages={117--128},
  year={2017}
}

@article{perscheid2017studying,
  title={Studying the advancement in debugging practice of professional software developers},
  author={Perscheid, Michael and Siegmund, Benjamin and Taeumel, Marcel and Hirschfeld, Robert},
  journal={Software Quality Journal},
  volume={25},
  number={1},
  pages={83--110},
  year={2017},
  publisher={Springer}
}

@article{Ko2006,
  title={An exploratory study of how developers seek, relate, and collect relevant information during software maintenance tasks},
  author={Ko, Amy J and Myers, Brad A and Coblenz, Michael J and Aung, Htet Htet},
  journal={IEEE Transactions on software engineering},
  volume={32},
  number={12},
  pages={971--987},
  year={2006},
  publisher={IEEE}
}

@article{beetlebox,
  title={BLAZE: Cross-language and cross-project bug localization via dynamic chunking and hard example learning},
  author={Chakraborty, Partha and Alfadel, Mahmoud and Nagappan, Meiyappan},
  journal={IEEE Transactions on Software Engineering},
  volume={51},
  number={8},
  pages={2254--2267},
  year={2025},
  publisher={IEEE}
}

@inproceedings{jones2002visualization,
  title={Visualization of test information to assist fault localization},
  author={Jones, James A and Harrold, Mary Jean and Stasko, John},
  booktitle={Proceedings of the 24th international conference on Software engineering},
  pages={467--477},
  year={2002}
}

@inproceedings{abreu2007accuracy,
  title={On the accuracy of spectrum-based fault localization},
  author={Abreu, Rui and Zoeteweij, Peter and Van Gemund, Arjan JC},
  booktitle={Testing: Academic and industrial conference practice and research techniques-MUTATION (TAICPART-MUTATION 2007)},
  pages={89--98},
  year={2007},
  organization={IEEE}
}

@article{wong2013dstar,
  title={The DStar method for effective software fault localization},
  author={Wong, W Eric and Debroy, Vidroha and Gao, Ruizhi and Li, Yihao},
  journal={IEEE Transactions on Reliability},
  volume={63},
  number={1},
  pages={290--308},
  year={2013},
  publisher={IEEE}
}

@article{sarhan2022survey,
  title={A survey of challenges in spectrum-based software fault localization},
  author={Sarhan, Qusay Idrees and Besz{\'e}des, {\'A}rp{\'a}d},
  journal={ieee access},
  volume={10},
  pages={10618--10639},
  year={2022},
  publisher={IEEE}
}

@article{Higor_2016_SBFL_Survey,
  title={Spectrum-based software fault localization: A survey of techniques, advances, and challenges},
  author={de Souza, Higor A and Chaim, Marcos L and Kon, Fabio},
  journal={arXiv preprint arXiv:1607.04347},
  year={2016}
}

@inproceedings{Papadakis_2012_Mutation,
  title={Using mutants to locate" unknown" faults},
  author={Papadakis, Mike and Le Traon, Yves},
  booktitle={2012 IEEE Fifth International Conference on Software Testing, Verification and Validation},
  pages={691--700},
  year={2012},
  organization={IEEE}
}

@inproceedings{Moon_2014_Mutation_MUSE,
  title={Ask the mutants: Mutating faulty programs for fault localization},
  author={Moon, Seokhyeon and Kim, Yunho and Kim, Moonzoo and Yoo, Shin},
  booktitle={2014 IEEE Seventh International Conference on Software Testing, Verification and Validation},
  pages={153--162},
  year={2014},
  organization={IEEE}
}

@article{Papadakis_2015_Metallaxis,
author = {Papadakis, Mike and Le Traon, Yves},
title = {Metallaxis-FL: mutation-based fault localization},
year = {2015},
issue_date = {August 2015},
publisher = {John Wiley and Sons Ltd.},
address = {GBR},
volume = {25},
number = {5–7},
issn = {0960-0833},
url = {https://doi.org/10.1002/stvr.1509},
doi = {10.1002/stvr.1509},
month = aug,
pages = {605–628},
numpages = {24}
}

@INPROCEEDINGS{Kim_2021_Mutation,
  author={Kim, Jinhan and An, Gabin and Feldt, Robert and Yoo, Shin},
  booktitle={2021 IEEE 32nd International Symposium on Software Reliability Engineering (ISSRE)}, 
  title={Ahead of Time Mutation Based Fault Localisation using Statistical Inference}, 
  year={2021},
  volume={},
  number={},
  pages={253-263},
  doi={10.1109/ISSRE52982.2021.00036}}

@inproceedings{Zhou_2012_IR_BugLocator,
author = {Zhou, Jian and Zhang, Hongyu and Lo, David},
title = {Where should the bugs be fixed? - more accurate information retrieval-based bug localization based on bug reports},
year = {2012},
isbn = {9781467310673},
publisher = {IEEE Press},
pages = {14–24},
numpages = {11},
location = {Zurich, Switzerland},
series = {ICSE '12}
}

@INPROCEEDINGS{Saha_2013_IR_BLUiR,
  author={Saha, Ripon K. and Lease, Matthew and Khurshid, Sarfraz and Perry, Dewayne E.},
  booktitle={2013 28th IEEE/ACM International Conference on Automated Software Engineering (ASE)}, 
  title={Improving bug localization using structured information retrieval}, 
  year={2013},
  volume={},
  number={},
  pages={345-355},
  doi={10.1109/ASE.2013.6693093}}

@inproceedings{Wen_2016_IR_Issues,
author = {Wen, Ming and Wu, Rongxin and Cheung, Shing-Chi},
title = {Locus: locating bugs from software changes},
year = {2016},
isbn = {9781450338455},
publisher = {Association for Computing Machinery},
address = {New York, NY, USA},
url = {https://doi.org/10.1145/2970276.2970359},
doi = {10.1145/2970276.2970359},
booktitle = {Proceedings of the 31st IEEE/ACM International Conference on Automated Software Engineering},
pages = {262–273},
numpages = {12},
location = {Singapore, Singapore},
series = {ASE '16}
}

@INPROCEEDINGS{Xuan_2014_LBFL_MULTRIC,
  author={Xuan, Jifeng and Monperrus, Martin},
  booktitle={2014 IEEE International Conference on Software Maintenance and Evolution}, 
  title={Learning to Combine Multiple Ranking Metrics for Fault Localization}, 
  year={2014},
  volume={},
  number={},
  pages={191-200},
  doi={10.1109/ICSME.2014.41}}

@article{Li_2017_LBFL_TraPT,
author = {Li, Xia and Zhang, Lingming},
title = {Transforming programs and tests in tandem for fault localization},
year = {2017},
issue_date = {October 2017},
publisher = {Association for Computing Machinery},
address = {New York, NY, USA},
volume = {1},
number = {OOPSLA},
url = {https://doi.org/10.1145/3133916},
doi = {10.1145/3133916},
month = oct,
articleno = {92},
numpages = {30}
}

@inproceedings{Li_2019_LBFL_DeepFL,
author = {Li, Xia and Li, Wei and Zhang, Yuqun and Zhang, Lingming},
title = {DeepFL: integrating multiple fault diagnosis dimensions for deep fault localization},
year = {2019},
isbn = {9781450362245},
publisher = {Association for Computing Machinery},
address = {New York, NY, USA},
url = {https://doi.org/10.1145/3293882.3330574},
doi = {10.1145/3293882.3330574},
pages = {169–180},
numpages = {12},
location = {Beijing, China},
series = {ISSTA 2019}
}

@inproceedings{Chen_2023_CL_SupConFL,
author = {Chen, Wei and Chen, Wu and Liu, Jiamou and Zhao, Kaiqi and Zhang, Mingyue},
title = {SupConFL: Fault Localization with Supervised Contrastive Learning},
year = {2023},
isbn = {9798400708947},
publisher = {Association for Computing Machinery},
address = {New York, NY, USA},
url = {https://doi.org/10.1145/3609437.3609441},
doi = {10.1145/3609437.3609441},
booktitle = {Proceedings of the 14th Asia-Pacific Symposium on Internetware},
pages = {44–54},
numpages = {11},
location = {Hangzhou, China},
series = {Internetware '23}
}

@INPROCEEDINGS {Li_2025_CL_FALCON,
author = { Li, Chun and Li, Hui and Li, Zhong and Pan, Minxue and Li, Xuandong },
booktitle = { 2025 IEEE/ACM 47th International Conference on Software Engineering (ICSE) },
title = {{ Enhancing Fault Localization in Industrial Software Systems via Contrastive Learning }},
year = {2025},
volume = {},
ISSN = {},
pages = {691-703},
doi = {10.1109/ICSE55347.2025.00009},
url = {https://doi.ieeecomputersociety.org/10.1109/ICSE55347.2025.00009},
publisher = {IEEE Computer Society},
address = {Los Alamitos, CA, USA},
month =May}

@inproceedings{Lou_2021_LBFL_GRACE,
author = {Lou, Yiling and Zhu, Qihao and Dong, Jinhao and Li, Xia and Sun, Zeyu and Hao, Dan and Zhang, Lu and Zhang, Lingming},
title = {Boosting coverage-based fault localization via graph-based representation learning},
year = {2021},
isbn = {9781450385626},
publisher = {Association for Computing Machinery},
address = {New York, NY, USA},
url = {https://doi.org/10.1145/3468264.3468580},
doi = {10.1145/3468264.3468580},
booktitle = {Proceedings of the 29th ACM Joint Meeting on European Software Engineering Conference and Symposium on the Foundations of Software Engineering},
pages = {664–676},
numpages = {13},
location = {Athens, Greece},
series = {ESEC/FSE 2021}
}

@misc{Sellami_2025_CLLLM,
      title={Contrastive Learning-Enhanced Large Language Models for Monolith-to-Microservice Decomposition}, 
      author={Khaled Sellami and Mohamed Aymen Saied},
      year={2025},
      eprint={2502.04604},
      archivePrefix={arXiv},
      primaryClass={cs.SE},
      url={https://arxiv.org/abs/2502.04604}, 
}

@misc{Meghwani_2025_Hard_Negatives,
      title={Hard Negative Mining for Domain-Specific Retrieval in Enterprise Systems}, 
      author={Hansa Meghwani and Amit Agarwal and Priyaranjan Pattnayak and Hitesh Laxmichand Patel and Srikant Panda},
      year={2025},
      eprint={2505.18366},
      archivePrefix={arXiv},
      primaryClass={cs.IR},
      url={https://arxiv.org/abs/2505.18366}, 
}

@inproceedings{Ciborowska_2022_LLM_FBL_BERT,
author = {Ciborowska, Agnieszka and Damevski, Kostadin},
title = {Fast changeset-based bug localization with BERT},
year = {2022},
isbn = {9781450392211},
publisher = {Association for Computing Machinery},
address = {New York, NY, USA},
url = {https://doi.org/10.1145/3510003.3510042},
doi = {10.1145/3510003.3510042},
booktitle = {Proceedings of the 44th International Conference on Software Engineering},
pages = {946–957},
numpages = {12},
location = {Pittsburgh, Pennsylvania},
series = {ICSE '22}
}

@inproceedings{Yang_2024_LLMAO,
author = {Yang, Aidan Z. H. and Le Goues, Claire and Martins, Ruben and Hellendoorn, Vincent},
title = {Large Language Models for Test-Free Fault Localization},
year = {2024},
isbn = {9798400702174},
publisher = {Association for Computing Machinery},
address = {New York, NY, USA},
url = {https://doi.org/10.1145/3597503.3623342},
doi = {10.1145/3597503.3623342},
booktitle = {Proceedings of the IEEE/ACM 46th International Conference on Software Engineering},
articleno = {17},
numpages = {12},
location = {Lisbon, Portugal},
series = {ICSE '24}
}

@misc{Chang_2025_LLM_BugCerberus,
      title={Bridging Bug Localization and Issue Fixing: A Hierarchical Localization Framework Leveraging Large Language Models}, 
      author={Jianming Chang and Xin Zhou and Lulu Wang and David Lo and Bixin Li},
      year={2025},
      eprint={2502.15292},
      archivePrefix={arXiv},
      primaryClass={cs.SE},
      url={https://arxiv.org/abs/2502.15292}, 
}

@article{Wong2016,
  title={A survey on software fault localization},
  author={Wong, W Eric and Gao, Ruizhi and Li, Yihao and Abreu, Rui and Wotawa, Franz},
  journal={IEEE Transactions on Software Engineering},
  volume={42},
  number={8},
  pages={707--740},
  year={2016},
  publisher={IEEE}
}

@article{DLFL_Survey,
author = {Niu, Feifei and Li, Chuanyi and Liu, Kui and Xia, Xin and Lo, David},
title = {When Deep Learning Meets Information Retrieval-based Bug Localization: A Survey},
year = {2025},
publisher = {Association for Computing Machinery},
address = {New York, NY, USA},
issn = {0360-0300},
url = {https://doi.org/10.1145/3734217},
doi = {10.1145/3734217},
journal = {ACM Comput. Surv.},
month = may
}

@inproceedings{codebert,
    title = "{C}ode{BERT}: A Pre-Trained Model for Programming and Natural Languages",
    author = "Feng, Zhangyin  and
      Guo, Daya  and
      Tang, Duyu  and
      Duan, Nan  and
      Feng, Xiaocheng  and
      Gong, Ming  and
      Shou, Linjun  and
      Qin, Bing  and
      Liu, Ting  and
      Jiang, Daxin  and
      Zhou, Ming",
    editor = "Cohn, Trevor  and
      He, Yulan  and
      Liu, Yang",
    booktitle = "Findings of the Association for Computational Linguistics: EMNLP 2020",
    month = nov,
    year = "2020",
    address = "Online",
    publisher = "Association for Computational Linguistics",
    url = "https://aclanthology.org/2020.findings-emnlp.139/",
    doi = "10.18653/v1/2020.findings-emnlp.139",
    pages = "1536--1547",
}

@article{Chakraborty_2024_RL_RLocator,
author = {Chakraborty, Partha and Alfadel, Mahmoud and Nagappan, Meiyappan},
title = {RLocator: Reinforcement Learning for Bug Localization},
year = {2024},
issue_date = {Oct. 2024},
publisher = {IEEE Press},
volume = {50},
number = {10},
issn = {0098-5589},
url = {https://doi.org/10.1109/TSE.2024.3452595},
doi = {10.1109/TSE.2024.3452595},
journal = {IEEE Trans. Softw. Eng.},
month = oct,
pages = {2695–2708},
numpages = {14}
}

@inproceedings{Gupta_2019_RL_Test_Prioritization,
author = {Gupta, Rahul and Kanade, Aditya and Shevade, Shirish},
title = {Deep reinforcement learning for syntactic error repair in student programs},
year = {2019},
isbn = {978-1-57735-809-1},
publisher = {AAAI Press},
url = {https://doi.org/10.1609/aaai.v33i01.3301930},
doi = {10.1609/aaai.v33i01.3301930},
booktitle = {Proceedings of the Thirty-Third AAAI Conference on Artificial Intelligence and Thirty-First Innovative Applications of Artificial Intelligence Conference and Ninth AAAI Symposium on Educational Advances in Artificial Intelligence},
articleno = {115},
numpages = {8},
location = {Honolulu, Hawaii, USA},
series = {AAAI'19/IAAI'19/EAAI'19}
}

@article{Chen_2024_RL_Test_Generation,
  title={Deep Reinforcement Learning-Based Automatic Test Case Generation for Hardware Verification},
  author={Chen, Jingyi and Yan, Lei and Wang, Shikai and Zheng, Wenxuan},
  journal={Journal of Artificial Intelligence General science (JAIGS) ISSN: 3006-4023},
  volume={6},
  number={1},
  pages={409--429},
  year={2024}
}

@article{Esnaashari_2021_RL_Test_Gen,
  title={Automation of software test data generation using genetic algorithm and reinforcement learning},
  author={Esnaashari, Mehdi and Damia, Amir Hossein},
  journal={Expert Systems with Applications},
  volume={183},
  pages={115446},
  year={2021},
  publisher={Elsevier}
}

@article{
Shojaee_2023_RL_Code_Generation,
title={Execution-based Code Generation using Deep Reinforcement Learning},
author={Parshin Shojaee and Aneesh Jain and Sindhu Tipirneni and Chandan K. Reddy},
journal={Transactions on Machine Learning Research},
issn={2835-8856},
year={2023},
url={https://openreview.net/forum?id=0XBuaxqEcG},
note={}
}

@inproceedings{Maruyama_2008_RL_Model_FL,
  title={Model-based fault localization in large-scale computing systems},
  author={Maruyama, Naoya and Matsuoka, Satoshi},
  booktitle={2008 IEEE International Symposium on Parallel and Distributed Processing},
  pages={1--12},
  year={2008},
  organization={IEEE}
}

@article{Liu_2005_RL_Model_SOBER,
  title={SOBER: statistical model-based bug localization},
  author={Liu, Chao and Yan, Xifeng and Fei, Long and Han, Jiawei and Midkiff, Samuel P},
  journal={ACM SIGSOFT Software Engineering Notes},
  volume={30},
  number={5},
  pages={286--295},
  year={2005},
  publisher={ACM New York, NY, USA}
}

@INPROCEEDINGS{Moran_2023_RL_SBFL4RL,
  author={Morán, Jesús and Bertolino, Antonia and De La Riva, Claudio and Tuya, Javier},
  booktitle={2023 IEEE International Conference On Artificial Intelligence Testing (AITest)}, 
  title={Fault Localization for Reinforcement Learning}, 
  year={2023},
  volume={},
  number={},
  pages={49-50},
  doi={10.1109/AITest58265.2023.00016}}

@article{Barto2019,
  author    = {Barto, Andrew and Mahadevan, Sridhar},
  title     = {Recent Advances in Hierarchical Reinforcement Learning},
  journal   = {Foundations and Trends in Machine Learning},
  year      = {2019},
  volume    = {14},
  number    = {1-2},
  pages     = {1-295},
  doi       = {10.1561/2200000071}
}

@inproceedings{
    swebench,
    title={{SWE}-bench: Can Language Models Resolve Real-world Github Issues?},
    author={Carlos E Jimenez and John Yang and Alexander Wettig and Shunyu Yao and Kexin Pei and Ofir Press and Karthik R Narasimhan},
    booktitle={The Twelfth International Conference on Learning Representations},
    year={2024},
    url={https://openreview.net/forum?id=VTF8yNQM66}
}

@inproceedings{nijkamp2023codegen,
  title={CodeGen: An Open Large Language Model for Code with Multi-Turn Program Synthesis},
  author={Erik Nijkamp and Bo Pang and Hiroaki Hayashi and Lifu Tu and Huan Wang and Yingbo Zhou and Silvio Savarese and Caiming Xiong},
  booktitle={The Eleventh International Conference on Learning Representations},
  year={2023},
  url={https://openreview.net/forum?id=iaYcJKpY2B_}
}

@inproceedings{Java_stack_traces,
author = {Medeiros, Marcos and Kulesza, Uira and Coelho, Roberta and Bonifacio, Rodrigo and Treude, Christoph and Barbosa, Eiji Adachi},
title = {The Impact Of Bug Localization Based on Crash Report Mining: A Developers' Perspective},
year = {2024},
isbn = {9798400705014},
publisher = {Association for Computing Machinery},
address = {New York, NY, USA},
url = {https://doi.org/10.1145/3639477.3639730},
doi = {10.1145/3639477.3639730},
booktitle = {Proceedings of the 46th International Conference on Software Engineering: Software Engineering in Practice},
pages = {13–24},
numpages = {12},
location = {Lisbon, Portugal},
series = {ICSE-SEIP '24}
}

@article{kaelbling1996reinforcement,
  title={Reinforcement learning: A survey},
  author={Kaelbling, Leslie Pack and Littman, Michael L and Moore, Andrew W},
  journal={Journal of artificial intelligence research},
  volume={4},
  pages={237--285},
  year={1996}
}

@article{wilcoxon1945,
  title={Individual comparisons by ranking methods},
  author={Wilcoxon, Frank},
  journal={Biometrics bulletin},
  volume={1},
  number={6},
  pages={80--83},
  year={1945},
  publisher={JSTOR}
}
